\documentclass[a4paper,12pt]{article}
\pdfoutput=1

\usepackage[utf8x]{inputenc}
\usepackage[T1]{fontenc}
\usepackage{lmodern}

\usepackage[lmargin=2.5cm,rmargin=2.5cm,tmargin=2.5cm,bmargin=3.5cm]{geometry}
\usepackage[small,font=it,labelfont=bf,width=1\textwidth]{caption}
\usepackage{amsmath,amssymb}
\usepackage{graphicx}
\usepackage[absolute]{textpos}
\usepackage{color}
\usepackage{booktabs}
\usepackage{xspace}
\usepackage{xcolor}
\usepackage{slashed}
\usepackage[normalem]{ulem}
\usepackage[bookmarks=true,linktocpage]{hyperref}
\usepackage{cite}
\usepackage{multirow}
\hypersetup{%
  pdftitle={Strong first-order phase transitions in the NMSSM --- a comprehensive survey},
  pdfauthor={Peter Athron, Csaba Balazs, Andrew Fowlie, Giancarlo Pozzo, Graham White, Yang Zhang},
  pdfkeywords={NMSSM, baryogenesis, electroweak},
  colorlinks=true,urlcolor=blue,linkcolor=blue,citecolor=blue,filecolor=blue}

\makeatletter
\g@addto@macro\bfseries{\boldmath}
\makeatother

\newcommand{\cosmo}{\texttt{Cosmo\-Transitions}\xspace}
\newcommand{\FS}{\texttt{Flexible\-SUSY}\xspace}
\newcommand{\SARAH}{\ensuremath{\texttt{SARAH}}\xspace}
\newcommand{\SARAHv}{\ensuremath{\texttt{SARAH-4.12.3}}\xspace}

\newcommand{\be}{\begin{equation}}
\newcommand{\ee}{\end{equation}}
\newcommand{\ba}{\begin{array}}
\newcommand{\ea}{\end{array}}
\newcommand{\bea}{\begin{eqnarray}}
\newcommand{\eea}{\end{eqnarray}}

\newcommand{\norm}[1]{\left\lVert#1\right\rVert}

\newcommand{\Huzr}{h_u}
\newcommand{\Hdzr}{h_d}

\newcommand{\Sr}{s}

\newcommand{\vu}{v_u}
\newcommand{\vd}{v_d}
\newcommand{\vs}{v_S}

\newcommand{\gev}{\ensuremath{\,\text{GeV}}}
\newcommand{\tev}{\ensuremath{\,\text{TeV}}}

\newcommand{\minima}{\mathfrak}

\newcommand{\msusy}{m_\text{SUSY}}
\newcommand{\mueff}{\mu_\text{eff}}

\newcommand{\MSbar}{\ensuremath{\overline{\text{MS}}}\xspace}

\DeclareMathOperator{\re}{Re}

\DeclareMathOperator{\sign}{sign}
\allowdisplaybreaks

\newcommand{\type}[1]{\textup{\texttt{Type-\allowbreak#1}}\xspace}
\newcommand{\TypeEWS}{\type{H-and-S}}
\newcommand{\TypeEW}{\type{Only-H}}
\newcommand{\TypeS}{\type{Only-S}}
\newcommand{\TypeSs}{\type{Only-S\allowbreak(maintain)}}
\newcommand{\TypeSf}{\type{Only-S\allowbreak(flip)}}

\newcommand{\gammaEW}{\ensuremath{\gamma_\text{EW}}\xspace}

\usepackage{siunitx} 
\usepackage{microtype} 
\usepackage{mathtools} 

\newcommand*\diff{\mathop{}\!\mathrm{d}}

\newcommand{\subs}[1]{\ensuremath{_\textup{#1}}} 
\newcommand{\sups}[1]{\ensuremath{^\textup{#1}}} 

\providecommand*{\eu}{\ensuremath{e}} 

\newcommand{\JB}{J_\text{B}}
\newcommand{\JF}{J_\text{F}}
\newcommand{\JBF}{J_\text{B/F}}

\newcommand{\reftab}[1]{Tab.~\ref{#1}}
\newcommand{\refsec}[1]{Sec.~\ref{#1}}
\newcommand{\refapp}[1]{App.~\ref{#1}}
\newcommand{\refcite}[1]{Ref.~\cite{#1}}
\newcommand{\reffig}[1]{Fig.~\ref{#1}}

\newcommand{\globalmin}{SM vacuum\xspace}

\title{\bf Strong first-order phase transitions in the NMSSM --- a comprehensive survey}

\author{Peter Athron$^a$, Csaba Balazs$^a$, Andrew Fowlie$^{a,b}$, Giancarlo Pozzo$^a$, \\
Graham White$^c$, Yang Zhang$^a$}

\date{}

\begin{document}
\maketitle
\thispagestyle{empty}
\begin{center}
  \it
  $^a$ARC Centre of Excellence for Particle Physics at the
  Terascale,\\ School of Physics and Astronomy, Monash University, Victoria 3800\\[0.5em]
  ${}^b$ Department of Physics and Institute of Theoretical Physics,\\ Nanjing Normal University, Nanjing, 210023, China\\[0.5em]
  ${}^c$ TRIUMF, 4004 Wesbrook Mall, Vancouver,\\
  British Columbia V6T 2A3, Canada
\end{center}
\begin{abstract}
Motivated by the fact that the Next-to-Minimal Supersymmetric Standard Model is one of the most plausible models that can accommodate electroweak baryogenesis, we analyze its phase structure by tracing the temperature dependence of the minima of the effective potential.
Our results reveal rich patterns of phase structure that end in the observed electroweak symmetry breaking vacuum.
We classify these patterns according to the first transition in their history and show the strong first-order phase transitions that may be possible in each type of pattern.
These could allow for the generation of the matter-antimatter asymmetry or potentially observable gravitational waves.
For a selection of benchmark points, we checked that the phase transitions completed and calculated the nucleation temperatures.
We furthermore present samples that feature strong first-order phase transitions from an extensive scan of the whole parameter space.
We highlight common features of our samples, including the fact that the Standard Model like Higgs is often not the lightest Higgs in the model.
\end{abstract}

\begin{textblock*}{7em}(\textwidth,1cm)
\noindent\footnotesize
CoEPP-MN-19-03 \\
\end{textblock*}

\clearpage

\section{Introduction}\label{sec:introduction}
One of the enduring problems in modern physics is the origin of the baryon asymmetry of the Universe (BAU)~\cite{Morrissey:2012db,Cline:2006ts,White:2016bo}. This asymmetry cannot be an initial condition in any cosmology that includes inflation, as that would wash out any initial asymmetry.\footnote{For an exception to this rule of thumb see \refcite{Krnjaic:2016ycc}.
} Therefore baryon asymmetry must be produced; however, as yet there is no experimental confirmation of any production mechanism.
Any mechanism that produces the BAU must satisfy three criteria~\cite{Sakharov:1967dj}:
\begin{enumerate}
    \item charge (C) and charge-parity (CP) violation,
    \item baryon number (B) violation, and
    \item departure from equilibrium.
\end{enumerate}
The Standard Model (SM) has the ingredients to satisfy all three criteria: there is a CP violating phase in the CKM matrix, B is violated through sphalerons which are unsuppressed at high temperature and there could be departures from equilibrium following two phase transitions (PTs) that occur in the SM vacuum as it cools --- the electroweak (EW) and the QCD transition. Quantitatively, however, the CP violating phase in the CKM matrix is far too feeble to produce enough baryon asymmetry. Furthermore the two transitions that occur in the SM at high temperature are both crossover transitions rather than first-order phase transitions (FOPTs) and therefore do not provide a large enough departure from equilibrium (see e.g., \refcite{DOnofrio:2015gop}). As such one has to look beyond the SM for explanations.

While the origin of the baryon asymmetry is a mystery, its measurement is on a firm foundation. During big bang nucleosynthesis, the baryon asymmetry is an input to the set of Boltzmann equations that govern the production of primordial light elements. Since we can measure some of these primordial abundances (deuterium in particular) with high accuracy, this constrains the baryon asymmetry\footnote{We convert measurements of the photon-baryon ratio to $Y_B$ by \refcite{Cline:2006ts}
\begin{equation}
Y_B \equiv \frac{n_B}{s} \approx \frac{1}{7.04} \frac{n_B}{n_\gamma}.
\end{equation}
}
to be~\cite{Patrignani:2016xqp}
\begin{equation}
    Y_B \equiv \frac{n_B}{s} =
    8.2\text{ -- }9.4 \times 10^{-11} ~(95\%~\text{CL}).
\end{equation}
Furthermore the baryon asymmetry produces acoustic oscillations in the power spectrum of the  cosmic microwave background (CMB)~\cite{Ade:2015xua}. Observing these oscillations gives an even tighter bound on the BAU,
\begin{equation}
    Y_B = 8.65 \pm 0.09 \times 10^{-11}.
\end{equation}
The fact that there is a concordance between these two unrelated measurement approaches is a triumph of cosmology. Along with dark matter and inflation, the origin of the BAU is a powerful cosmological argument for physics beyond the SM.

Electroweak baryogenesis is a minimal and natural explanation for the origin of the baryon asymmetry in the Universe~\cite{Trodden:1998ym, Cline:2008hr,Cline:2009sn,Borah:2012pu,Cline:2013bln,Konstandin:2013caa,Kozaczuk:2014kva,Profumo:2014opa,Curtin:2014jma,Huang:2015bta, Inoue:2015pza,Katz:2015uja,Fuyuto:2015ida,Huang:2015izx,Kobakhidze:2015xlz,Huang:2016odd,Kotwal:2016tex,White:2016bo,Vaskonen:2016yiu,Balazs:2016yvi, Beniwal:2017eik,Kurup:2017dzf,Akula:2017yfr,Chiang:2017nmu,Cao:2017oez,Ramsey-Musolf:2017tgh,Huang:2017kzu,deVries:2017ncy, Niemi:2018asa, Modak:2018csw,Carena:2018cjh,Chala:2018opy, Zhou:2018zli,Alves:2018jsw,YaserAyazi:2019caf,Mohamadnejad:2019vzg}. It utilizes the electroweak phase transition (EWPT) which is known to have occurred in our cosmic history providing the reheating temperature was not unnaturally small. Although this transition is a crossover in the SM, its character may be modified by the introduction of new weak scale bosons such that the transition becomes a strongly FOPT (SFOPT) and proceeds by bubble nucleation.  Such a phenomenon is all the more interesting because it might directly be probed by future gravitational wave detectors~\cite{Witten:1984rs, 1986MNRAS.218..629H, Krauss:1991qu, Kosowsky:1991ua, Kosowsky:1992rz, Kamionkowski:1993fg}.

This mechanism can be in principle realized within supersymmetry. In the Minimal Supersymmetric Standard Model (MSSM) a barrier between the EW symmetric and broken vacuums arises from thermal corrections from stops; however, one requires light stops to catalyze the PT such that it is sufficiently strongly first order~\cite{Lee:2004we,Balazs:2004ae}. This is all but ruled out by LHC constraints on stop masses~\cite{Liebler:2015ddv}. Much more attractive is the possibility of the Next-to-Minimal Supersymmetric Standard Model (NMSSM)~\cite{Maniatis:2009re,Ellwanger:2009dp} where a light singlet scalar can catalyze a strongly first order EWPT~\cite{Profumo:2014opa, Kotwal:2016tex, Li:2019tfd}. Unlike the stop which catalyzes the PT through thermal effects, the singlet can change the potential such that there is a barrier even at zero temperature.

Electroweak baryogenesis was recently considered within the NMSSM~\cite{Bian:2017wfv,Huber:2006wf,Kozaczuk:2014kva,Balazs:2013cia,Cheung:2012pg,Huang:2014ifa} and it was found that the baryon asymmetry can vary by an order of magnitude depending on whether the singlet acquires a vacuum expectation value (VEV) before or during the EWPT (with a simultaneous transition providing more efficient baryon production)~\cite{Balazs:2013cia}. Furthermore the baryon yield is proportional to the maximal variation of the ratio of the two Higgs VEVs, $\Delta \beta$, and it was shown in \refcite{Kozaczuk:2014kva,Demidov:2016wcv,Bi:2015qva,Carena:2011jy,Menon:2004wv,Akula:2017yfr} that $\Delta \beta$ can be an order of magnitude larger in the NMSSM compared to the MSSM.

In this work we explore the plausibility of EW baryogenesis within the NMSSM, focusing on the PT and leaving CP violation to future work (see~\cite{Bell:2019mbn,deVries:2018tgs,Chen:2017com,Guo:2016ixx,Chao:2015uoa,Cline:2017qpe,Cline:2012hg,Cline:2011mm,Bian:2017wfv,Carena:2018vpt,Cline:1997vk,Grzadkowski:2018nbc,Ellis:2019flb,Huang:2018aja} for various approaches to generating CP violation). We consider the case where the superpartners are all heavy enough to have their thermal contributions Boltzmann suppressed during the transition. Thus we can match our model to a two Higgs doublet model plus a singlet (THDMS).  We sample the parameter space to find points with an EW SFOPT. For such points, we investigate the phase structure, that is the evolution of the minima of the effective potential as the Universe cools.  This investigation includes determining whether the singlet acquires a VEV during or before the EWPT and it also involves calculating the strength of the PT.

As we focus on the third Sakharov condition (a departure from thermal equilibrium),
we do not consider explicit
or spontaneous CP violation in the Higgs sector. We instead assume that CP violation enters
the Higgs sector radiatively, though remain agnostic about the exact source of CP violation and
do not examine constraints on complex phases (such as electric dipole moments).
This simplification allows us to
focus only on PTs between the ground states of CP-even fields,
easing the numerical problem of finding vacua of
a multifield scalar potential.

The structure of this paper is as follows.  In \refsec{sec:models} we introduce the NMSSM and the THDMS, fixing the notation we will use in the
paper. Following this, in \refsec{sec:potential} we describe
the radiative and finite temperature corrections that we include in
our analysis. Then in \refsec{sec:FOPT} we outline the procedure
we use to determine if a point in the parameter space has a FOPT or not,
and if so calculate the critical temperature and transition strength.
The results of our scan are presented in \refsec{sec:results} and
finally our conclusions are given in \refsec{sec:conclusions}.

\section{NMSSM}\label{sec:models}

The NMSSM extends the MSSM particle content by adding one singlet superfield, $\hat{S}$.  Here we work in the $\mathbb{Z}_3$ symmetric NMSSM where the $\mu$-term of the MSSM is forbidden and instead an effective $\mu$-term, $\mueff = \lambda \langle S \rangle$, is generated when the singlet develops a VEV, thus solving the $\mu$-problem of the MSSM. The superpotential is given by
\be
\mathcal{W}_{\textrm{NMSSM}} = (Y_u)_{ij} \, \hat{Q}_i\cdot\hat{H}_u \, \hat{u}^c_j   + (Y_d)_{ij} \, \hat{Q}_i\cdot\hat{H}_{d} \, \hat{d}^c_j +  (Y_e)_{ij} \, \hat{L}_i\cdot\hat{H}_{d} \, \hat{e}^c_j, + \lambda \, \hat{S} \, \hat{H}_u \cdot\hat{H}_d + \frac13\kappa \, \hat{S}^3,
\label{Eq:superpotential}
\ee
where a hat is used for superfields, $i,j \in \{1,2,3\}$ are family indices, and we have introduced the $SU(2)_L$ dot product, $A \cdot B = A^1 B^2 - A^2 B^1$. The discrete $\mathbb{Z}_3$ symmetry is spontaneously broken when the Higgs fields or singlet obtain a VEV. We assume that following the strategies of \refcite{Abel:1996cr,Panagiotakopoulos:1998yw,Panagiotakopoulos:1999ah} domain wall problems can in principle be avoided without impacting any phenomenology.

Under the SM gauge group $G_\text{SM} = SU(3)_C\times SU(2)_L \times U(1)_Y$ the superfields transform as
\be
\begin{aligned}
\hat{Q}   &:\textstyle (\mathbf{3}, \mathbf{2}, \frac{1}{6}),
&\hat{u}^c&:\textstyle (\mathbf{\bar{3}},\mathbf{1},-\frac{2}{3}),
&\hat{d}^c&:\textstyle (\mathbf{\bar{3}},\mathbf{1}, \frac{1}{3}),
&\hat{L}  &:\textstyle (\mathbf{1}, \mathbf{2},-\frac{1}{2}),
&\hat{e}^c&:\textstyle (\mathbf{1}, \mathbf{1}, 1),
\\
 \hat{H}_d&:\textstyle (\mathbf{1}, \mathbf{2},-\frac{1}{2}),
&\hat{H}_u&:\textstyle (\mathbf{1}, \mathbf{2}, \frac{1}{2}),
  &\hat{S}&:\textstyle (\mathbf{1}, \mathbf{1}, 0)
\end{aligned}
\label{Eq:superfields}
\ee
where the first two entries inside the
parentheses give the representation under $SU(3)_C$ and
$SU(2)_L$, respectively, while the third entry gives the $U(1)_Y$
hypercharges without GUT normalization.

There are three contributions to the tree-level Higgs potential of the NMSSM:
\be
V_\textrm{NMSSM} = V_F + V_D + V_\textrm{soft} .
\label{Eq:NMSSMpotential}
\ee
Here the $F$- and $D$-term contributions are
\begin{align}
V_F &= \left
|\lambda S \right|^2 (|H_u|^2 + |H_d|^2) + \left |\lambda H_u \cdot
H_d + \kappa S^2 \right |^2, \\
V_D &= \frac18(g^2 + g^{\prime
  2})(|H_u|^2 - |H_d|^2)^2 + \frac12 g^2|H_u^\dagger H_d|^2,
\end{align}
where $g$ and $g^\prime$ are respectively the $SU(2)_L$
and $U(1)_Y$ gauge couplings without GUT normalization.
Finally, the soft-breaking terms are
\begin{equation}
V_\textrm{soft} = m_{H_u}^2 |H_u|^2 + m_{H_d}^2 |H_d|^2 + m_{S}^2
|S|^2 + [\lambda A_\lambda S H_u \cdot H_d + \frac13 \kappa A_\kappa S^3 +
  \textrm{h.c.}].
\end{equation}
The couplings $\lambda$ and $\kappa$ and the corresponding trilinears, $A_\lambda$ and $A_\kappa$, are in general complex.
Three of the four phases, however, may be removed through field redefinitions of $H_u$, $H_d$ and $S$.
Since current LHC limits and the $125\gev$ Higgs mass measurements
require squarks and gluinos to be TeV-scale, the mass
spectrum of the NMSSM must contain a large hierarchy between the SM particles and colored sparticles.
Furthermore the states with the
largest couplings include both heavy sparticles and light SM particles,
i.e., stops and the top quark. Therefore higher-order corrections will
always include large logarithms since one cannot choose the
renormalization scale $Q$ to simultaneously minimize $\ln m_t / Q$ and
$\ln M_\text{SUSY} / Q$.  This makes it challenging to perform
precise calculations when working in the full theory. To improve the
precision of our calculations we will integrate out the heavy
superpartners and use an effective field theory (EFT) which contains
only the light states. This makes it possible run to $Q = m_t$ and
perform calculations in the EFT which are free from large
logarithms.

\subsection{Matching to the THDMS}
\label{sec:THDMS}

Since we want to consider scenarios in which all superpartners are too heavy to impact the PT, we match the
NMSSM to a two Higgs
doublet model plus a singlet (THDMS), which in this context is an
effective field theory of the full NMSSM theory valid below $M_\text{SUSY}$.\footnote{This is also the approach taken in Refs.~\cite{Kozaczuk:2014kva,Elliott:1993ex,Elliott:1993uc,Elliott:1993bs}.}

The tree-level potential of a $\mathbb{Z}_3$ symmetric THDMS model is
\be
\begin{aligned}
V\sups{tree}_\textrm{THDMS}
&=
\frac12 \lambda_1 |H_d|^4 + \frac12 \lambda_2 |H_u|^4 + (\lambda_3 + \lambda_4) |H_u|^2 |H_d|^2 - \lambda_4 |H_u^\dagger H_d|^2\\
&+ \lambda_5 |S|^2|H_d|^2 + \lambda_6 |S|^2 |H_u|^2 + (\lambda_7 S^{*2}H_d \cdot H_u + \textrm{h.c.}) + \lambda_8 |S|^4 \\
&+ m_1^2 |H_d|^2 + m_2^2|H_u|^2 + m_3^2 |S|^2 - (m_4 S H_d \cdot H_u + \textrm{h.c.}) - \frac13(m_5 S^3 + \textrm{h.c.}),
\end{aligned}
\label{Eq:THDMS_tree_Pot}
\ee
where the couplings $\lambda_7$, $m_4$ and $m_5$ may be complex. Two of the three phases, however, may be removed by redefinitions of $H_u$, $H_d$ and $S$,
leaving a single complex phase, as in the NMSSM.
In~\eqref{Eq:THDMS_tree_Pot} we follow the conventions in \refcite{Elliott:1993ex,Elliott:1993uc,Elliott:1993bs,Kozaczuk:2014kva}; in particular the $|H_u|^2 |H_d|^2$ coefficient is $\lambda_3 + \lambda_4$.
We match the NMSSM to the THDMS at the scale $M\subs{SUSY}$ by identifying the tree-level conditions
\begin{align}
\label{eq:MatchingConditions}
\begin{gathered}
\lambda_1 = \frac{1}{4} \left({g'}^2+g^2\right),
\quad
\lambda_2 = \frac{1}{4} \left({g'}^2+g^2\right) + \Delta \lambda_2,
\quad
\lambda_3=\frac{1}{4} \left(g^2-{g'}^2\right),
\\
\lambda_4=\frac{1}{2} \left(2 |\lambda|^2-g^2\right),
\quad
\lambda_5=\lambda_6 =|\lambda|^2,
\quad
\lambda_7=-\lambda \kappa^*,
\quad
\lambda_8=|\kappa|^2,
\\
m_1^2 = m_{H_d}^2,
\quad
m_2^2 = m_{H_u}^2,
\quad
m_3^2 = m_{S}^2,
\quad
m_4 = A_\lambda \lambda,
\quad
m_5 = -A_\kappa \kappa.
\end{gathered}
\end{align}
We furthermore included a dominant one-loop threshold correction to the matching for $\lambda_2$,
\begin{equation}\label{eq:OneLoopThreshold}
\Delta \lambda_2
=
\frac{3 y_t^4 A_t^2}{8 \pi^2 M\subs{SUSY}^2}
\left(1-\frac{A_t^2}{12 M\subs{SUSY}^2} \right).
\end{equation}

Although we stated the potential and matching conditions for $\lambda_7$, $m_4$ and $m_5$ without
loss of generality, we later consider only real, CP conserving parameters.
As discussed in \refsec{sec:introduction} we assume that
the CP violation demanded by Sakharov's first condition originates in a different
sector of the NMSSM, e.g., the squark sector. Although CP violation must enter the Higgs sector through loops, since we only consider the dominant one-loop corrections in the matching, CP violating phases that may appear outside of the Higgs sector do not enter our calculation. At higher orders, however, we would be forced to consider complex parameters and consequently (as later discussed) PTs
involving CP-odd fields. An examination of the potential impact this could have is left for future study. Since we match the NMSSM to a THDMS, our results are also applicable to a subspace of the THDMS, which is well-motivated even in the absence of supersymmetry.

\section{Effective potential}
\label{sec:potential}

\subsection{Effective potential at zero temperature}

In the $R_\xi$ gauge the one-loop corrections to the potential, $\Delta V$, are given by~\cite{Patel:2011th}
\begin{equation}\label{eq:1LoopCorrection}
\begin{split}
\Delta V = \frac{1}{64 \pi^2} \Bigg( & \sum_h n_h m_h^4(\xi) \left[\ln\left( \frac{m_h^2(\xi)}{Q^2}\right) - 3/2\right]\\
+ & \sum_V n_V m_V^4 \left[\ln\left(\frac{m_V^2}{Q^2}\right) - 5/6\right]\\
- & \sum_V \tfrac13 n_V (\xi m_V^2)^2 \left[\ln\left(\frac{\xi m_V^2}{Q^2}\right) - 3/2\right]\\
- & \sum_f n_f m_f^4 \left[\ln\left(\frac{m_f^2}{Q^2}\right) - 3/2\right]\Bigg).
\end{split}
\end{equation}
where $Q$ is the renormalization scale, $m_i$ are field dependent \MSbar masses and the $n_i$ are the numbers of degrees of freedom for field $i$. The first term sums fluctuations of scalar fields, which at the EW breaking minimum can be separated into physical Higgs bosons and Goldstone bosons, the second term sums transverse and longitudinal massive gauge bosons, the third one scalar gauge boson fluctuations, and the final one fermions.

We neglect contributions to the vacuum energy. The numbers of degree of freedom for the particles that we include are
\begin{align}\label{eq:field_dofs}
n_{h^0_i} &= n_{A^0_i}  = n_{H^+_i} = n_{H^-_i} = 1,\\
n_{W^+} &=n_{W^-}=n_{Z} = 3,\\
n_t &= n_b =12,\, n_{\tau} = 4
\end{align}
for the real scalar, vector and Dirac fermion fields in our model, where $A^0_i$, $H ^+_i$ and $H^-_i$ include the physical Higgs states and the Goldstone bosons.

At zero temperature, the minimum of the one-loop potential lies at non-zero values for the Higgs fields, which we refer to as VEVs, and assume may always be written as
\begin{align}
\langle H_u\rangle = \frac{1}{\sqrt{2}} \begin{pmatrix} 0 \\ v_u \end{pmatrix}, \quad \quad
\langle H_d\rangle = \frac{1}{\sqrt{2}} \begin{pmatrix} v_d \\ 0 \end{pmatrix}, \quad \quad
\langle S\rangle = \frac{1}{\sqrt{2}} v_S,
\end{align}
where $v_u$, $v_d$ and $v_S$ are real, i.e., we do not consider charge or CP breaking VEVs.\footnote{Spontaneous charge and CP violation are impossible at tree-level in our THDMS model with NMSSM matching conditions~\cite{Romao:1986jy}. See, however, \refcite{Ferreira:2019iqb} for a recent discussion of this issue in a general THDMS model.} As we assume that the VEVs are CP conserving, a tadpole condition forces CP violating phases in the potential to vanish. 

 To construct the field dependent masses appearing in~\eqref{eq:1LoopCorrection}, we consider the potential as a function of the fields corresponding to the VEVs, i.e., we consider the $h_u$, $h_d$ and $s$ components of the fields,
\begin{align}
H_u = \begin{pmatrix} H_u^+\\ \tfrac{1}{\sqrt{2}} \left(h_u + i a_u\right) \end{pmatrix}, \quad \quad
H_d = \begin{pmatrix} \tfrac{1}{\sqrt{2}}  \left(h_d + i a_d\right) \\ H_d^- \end{pmatrix}, \quad \quad
S = \frac{1}{\sqrt{2}} \left(s + i \sigma\right),
\end{align}
where $h_u$, $h_d$ and $s$ are real. The field dependent masses are functions of $h_u$, $h_d$ and $s$. In principle, we could consider variation of the charged and CP-odd fields which cannot all be eliminated by gauge fixing. However, because we consider PTs only between charge and CP conserving vacua, we set charged and CP-odd Higgs fields to zero in the field dependent masses.  The expressions for the field dependent masses are given in \refapp{sec:Field_Dependent_Masses}.

The effective potential also contains explicit dependence on the gauge parameter $\xi$. The physical, gauge-independent content of the effective potential may be found through Nielsen identities~\cite{Nielsen:1975fs}, which express the fact that at extrema, $\minima{h}$, the gauge dependence of the effective potential vanishes, since
\begin{equation}
\frac{\partial V_\text{eff}(h, \xi)}{\partial \xi} \propto \frac{\partial V_\text{eff}(h, \xi)}{\partial h},
\end{equation}
and thus
\begin{equation}\label{eq:extrema_gauge_independent}
\frac{\diff V_\text{eff}(\minima{h}, \xi)}{\diff \xi} = \frac{\partial V_\text{eff}(\minima{h}, \xi)}{\partial \xi} + \frac{\partial \minima{h}}{\partial \xi}\frac{\partial V_\text{eff}(\minima{h}, \xi)}{\partial h} = 0.
\end{equation}
The location of the extrema, however, are gauge dependent, i.e., $\partial \minima{h} / \partial \xi \neq 0$. See e.g., \refcite{Patel:2011th,DiLuzio:2014bua} for further discussion of this issue. We work in the $\xi = 1$ (Feynman) gauge. The effective potential furthermore depends on a choice of renormalization scale, which could in fact have greater impact than gauge ambiguities~\cite{Laine:2017hdk}.

\subsection{Effective potential at finite temperature}\label{sec:FiniteTempEffPot}

To describe the conditions of the early Universe we need to take into
account temperature corrections. We calculate one-loop finite temperature corrections including daisy terms using the Arnold-Espinosa method~\cite{Arnold:1992rz} in the $\xi = 1$ (Feynman) gauge. The effective potential can be written as a sum of zero temperature and finite temperature pieces
\begin{equation}\label{Eq:FullPotential}
    V_\text{eff} = V^\text{tree}_\text{THDMS} +  \Delta V_\text{THDMS} + \Delta V_T + V_\text{daisy}.
\end{equation}
The one-loop thermal corrections in the $R_\xi$ gauge are~\cite{Patel:2011th}
\begin{equation}\label{eq:thermal_one_loop}
\begin{split}
\Delta V_T = \frac{T^4}{2 \pi^2} \Bigg[&
\sum_h n_h \JB\left(\frac{m_h^2(\xi)}{T^2}\right)
+ \sum_V n_V \JB\left(\frac{m_V^2}{T^2}\right)\\
- & \sum_V \tfrac13 n_V \JB\left(\frac{\xi m_V^2}{T^2}\right)
+ \sum_f n_f \JF\left(\frac{m_f^2}{T^2}\right)\Bigg] ,
\end{split}
\end{equation}
where
the field dependent masses are the same as those appearing in~\eqref{eq:1LoopCorrection} in the previous section, and the expressions for them are given in \refapp{sec:Field_Dependent_Masses}.  The degrees of freedom, $n$, are as in~\eqref{eq:field_dofs}; we again neglect contributions to the vacuum energy and the thermal functions are
\begin{equation}
\JBF(y^2) =
\pm \re
\int_0^{\infty}
		x^2 \ln
		\left(
			1 \mp \eu^{-\sqrt{x^2 + y^2}}
		\right)
\diff{x} .
\label{Eq:JBJF}
\end{equation}
Here the upper/lower signs are for bosons/fermions. For $m^2 \gg T^2$ the thermal functions are exponentially suppressed by a Boltzmann factor. This ensures that the massive supersymmetric particles that we integrated out do not impact the finite temperature corrections.

The daisy terms are
\begin{equation}
V \subs{daisy} = -\frac{T}{12 \pi} \left(
\sum_h n_h \left[ \left(\bar m_h^2\right)^{\frac32} - \left(m^2_h\right)^{\frac32} \right] + \sum_V \tfrac13 n_V \left[\left(\bar m_V^2\right)^{\frac32} - \left(m^2_V\right)^{\frac32} \right]
\right),
\end{equation}
where we sum over the Higgs fields (including Goldstone bosons) and massive gauge bosons, and $\bar m^2$ are field dependent
mass eigenvalues that include Debye corrections to the tree-level masses in the mass matrices.
The Debye corrections add additional $T$ dependent terms of the form $c_{\Phi} T^2 |\Phi|^2$ for all complex scalar gauge eigenstates and $c_A T^2 A_\mu A^\mu$ for all gauge bosons associated with the original gauge symmetries before EWSB.  For the THDMS we find,
\begin{align}
c_{H_u} &= \tfrac{1}{48} \left(3 {g'}^2+ 9 g^2 + 12 y_t^2 + 12 \lambda_2 + 8 \lambda_3 + 4 \lambda_4 + 4 \lambda_6 \right),\label{Eq:DebeyeCoeffs_Hu}\\
c_{H_d} &= \tfrac{1}{48} \left(3 {g'}^2+ 9 g^2 + 12 y_b^2 + 4 y_\tau^2 + 12 \lambda_1 + 8 \lambda_3 + 4 \lambda_4 + 4 \lambda_5 \right),\label{Eq:DebeyeCoeffs_Hd}\\
c_S &= \tfrac{1}{48} \left( 8 \lambda_5 + 8 \lambda_6 + 16 \lambda_8 \right),\label{Eq:DebeyeCoeffs_S}\\
c_{W_{1,2,3}} &= 2 g^2,\label{Eq:DebeyeCoeffs_W}\\
c_B &= 2 {g'}^2,\label{Eq:DebeyeCoeffs_B}
\end{align}
where the couplings ${g'}$, $g$, $y_t$, $y_b$ and $y_\tau$ are as in~\eqref{eq:GaugeBosonsTopBottomTauMasses}. The corrections for the gauge bosons are in the gauge basis before symmetry breaking and every component of a gauge representation receives the same Debye correction. The scalar coefficients are gauge independent, as they originate from a high-temperature expansion of~\eqref{eq:thermal_one_loop}, in which the dependence on $\xi$ cancels,
\begin{equation}
c_{ij} = \left.\frac{1}{T^2}\frac{\partial^2 \Delta V_T}{\partial \phi_i \partial \phi_j}\right|_{T^2 \gg m^2}.
\end{equation}
The coefficents for the gauge bosons are the same as those of the two-Higgs doublet model, which can be found in the literature \cite{Basler:2016obg}.  We cross-checked our results in~\eqref{Eq:DebeyeCoeffs_Hu} -- \eqref{Eq:DebeyeCoeffs_B} against general expressions in \refcite{Basler:2018cwe}. Thus we have described the full finite temperature potential, which is a function of the fields $h_u$, $h_d$ and $s$ and the temperature, $T$.

\section{First-order phase transitions}
\label{sec:FOPT}

Having constructed the finite temperature effective potential, we investigated whether there was a FOPT in which the vacuum of the potential changed abruptly as the Universe cooled. For such a transition to occur, the potential must exhibit two minima separated by a barrier. The temperature at which the two minima are exactly degenerate is known as the critical temperature. That is, at the critical temperature, $T_C$, there are minima such that
\begin{equation}\label{eq:TcDefinition}
V_\text{eff}(\minima{h}_u, \minima{h}_d, \minima{s}, T_C) = V_\text{eff}(\minima{h}^\prime_u, \minima{h}^\prime_d, \minima{s}^\prime, T_C)
\end{equation}
where caligraphic fonts, $\minima{h}_u$ etc, indicate a minimum of the scalar potential, i.e.,
\begin{equation}
\partial_{h_u} V_\text{eff}(\minima{h}_u, \minima{h}_d, \minima{s}) =
\partial_{h_d} V_\text{eff}(\minima{h}_u, \minima{h}_d, \minima{s}) =
\partial_s V_\text{eff}(\minima{h}_u, \minima{h}_d, \minima{s}) = 0.
\end{equation}
Below the critical temperature, the potential develops a minimum that is deeper than the other minima.  The system may tunnel through the barrier to the new vacuum state with the lower minimum~\cite{Coleman:1977py,Callan:1977pt,Linde:1980tt}.  As discussed below, however, the transition might not complete.

We developed a C++ program, \texttt{PhaseTracer}, to map the temperature dependence of the minima of the effective potential and to find potential PTs between them.  It enhances the algorithm that was developed in \cosmo~\cite{Wainwright:CosmoTransition} to map out the phase structure, and to find out possible PTs between every phase. The numerical method coded in \texttt{PhaseTracer} is briefly described in \refapp{sec:pt_methods}. This method is different from the one applied in the code \texttt{BSMPT} ~\cite{Basler:2018cwe} and previous works on SFOPTs in the NMSSM~\cite{Huang:2014ifa}, which may only find a single PT between the EW symmetric vacuum and the observed EWSB vacuum.  Our method is able to map out a more complicated phase structure and find multiple PTs in it.  Of equal importance, by analyzing the phase structure obtained by \texttt{PhaseTracer}, we confirmed that not all potential tunnelings actually take place in the early Universe.  This may happen because the tunneling rate is too slow or because the PT is located on a branch of the phase structure that the system never utilizes because it evolved in a different direction.

To exhibit spontaneous EWSB as the Universe cooled, the vacuum of the finite temperature effective potential~\eqref{Eq:FullPotential} should respect EW symmetry at high temperature, which is $1\tev$ in this work, and should violate it at zero temperature. Thus at high temperature the global minimum should lie at the origin, $h_u = 0$ and $h_d = 0$, and at zero temperature the deepest minimum should lie at the observed EWSB VEV.  We can use this information to fix the boundaries of the phase structure by finding all minima of the potential at $T=0$ and $T=1\tev$ and checking that spontaneous symmetry breaking occurs.  Starting from $T=0$ then we can use \texttt{PhaseTracer} to find all possible PTs.

The strength of such a transition is described by an order parameter. For baryogenesis, we consider the order parameter
\begin{equation}\label{eq:gamma}
\gammaEW \equiv \frac{\sqrt{
(\minima{h}_u - \minima{h}^\prime_u)^2 + (\minima{h}_d - \minima{h}^\prime_d)^2}}{T_C}.
\end{equation}
The singlet VEV is not included here because it does not affect EW sphalerons.
Order parameters of about $\gammaEW \gtrsim 1$ are considered strong and could catalyze baryogenesis.

The Nielsen identities in~\eqref{eq:extrema_gauge_independent} imply that the critical temperature is gauge independent, since the effective potential is gauge independent at extrema. Our one-loop truncation of the effective potential, however, means that it is gauge independent only at the tree-level extrema.
Thus the critical temperature, which we find from the effective potential at the one-loop minima, is gauge dependent. See \refcite{Patel:2011th} for further discussion and a procedure that may enforce gauge independence. The location of the minima, furthermore, and thus the order parameter, always depend on the gauge parameter $\xi$.

A first order transition occurs through bubble nucleation and there is a finite probability per unit time and volume for tunneling to a new phase. The new phase dominates once the following condition is satisfied~\cite{McLerran:1990zh, Dine:1991ck},
\begin{equation}\label{eq:define_TN}
 \frac{S_E(T_N)}{T_N} \simeq 140,
\end{equation}
where $S_E$ stands for the Euclidean bubble action, and $T_N$ is the so-called nucleation temperature. If there is no solution, we conclude that the transition cannot complete. During the scan, we identify all possible PTs without checking whether they successfully nucleate. After classifying phase structures, we check nucleation temperatures for a subset of our samples using \cosmo~\cite{Wainwright:CosmoTransition}.

\section{Results}\label{sec:results}

\subsection{Parameter space, constraints and sampling strategy}

To explore all possible PTs in the NMSSM, including
strong EWPTs, we sampled the parameter space
of the model within the ranges shown in \reftab{Tab:scan}. The first
four parameters, $\lambda$, $\kappa$, $A_\lambda$ and $A_\kappa$ are
from the tree-level NMSSM potential and enter the matching conditions
at tree-level~\eqref{eq:MatchingConditions}, while the fifth
parameter, the stop trilinear $A_t$ enters at the one-loop level~\eqref{eq:OneLoopThreshold}. These parameters are all defined at the
matching scale $\msusy$ which we also take as an input and represents
the geometric mean of the left and right soft SUSY breaking masses of the stops, which have been integrated out, i.e.
\begin{equation}
\msusy = \sqrt{m_{\tilde{t}_L} m_{\tilde{t}_R}} .
\end{equation}
The final two parameters are the ratio of the Higgs VEVs $\tan\beta \equiv
v_u / v_d$ and the singlet VEV, $v_S$, which are defined at the top
quark mass, $m_t = 173.1\gev$.  Therefore our model has eight free
parameters.

From these inputs the parameters of the THDMS at $m_t$ are obtained
using \texttt{\FS-2.1.0}~\cite{Athron:2014yba,Athron:2017fvs}, coupled
with\footnote{Internally \FS also uses some numerical routines from
  \texttt{SOFTSUSY}~\cite{Allanach:2001kg,Allanach:2013kza}.}
\SARAHv~\cite{Staub:2009bi,Staub:2010jh,Staub:2012pb,Staub:2013tta},
which implements the matching and running procedure described in
\refsec{sec:THDMS}, with~\eqref{eq:MatchingConditions} specified as a
boundary condition in the \FS model file.\footnote{The \SARAH and \FS
  model files we wrote for this are provided as supplementary material
  to our arXiv submission.}  Since all running and effective potential
calculations are performed in the THDMS it is not necessary to specify
any further soft-breaking masses in the NMSSM. Because the quartic
coupling $\lambda$ can always be made positive through field
redefinitions, we do not consider negative values for it, but we do
consider both negative and positive values for the soft
trilinears, $\kappa$ and $v_S$. Lastly, as discussed earlier, for
self-consistency we only consider real parameters.

The field dependent masses which enter the one-loop corrections to
potential are calculated with \FS, and the thermal functions are
evaluated using the implementation described in
\refcite{Fowlie:2018eiu}. We use \texttt{PhaseTracer} to find the
phases and critical temperatures by exploring the finite temperature
potential between $T=0$ and $T= 1$\tev, as described in
\refsec{sec:FOPT}. Since this involves varying the field values that
enter the field-dependent masses, in principle it is possible that
this could re-introduce large logarithms and lead to perturbativity
problems, therefore we do not consider VEVs greater than $1.6\tev$.
In practice in all our results the VEVs are significantly
smaller than this, and are less than $300\gev$ in all but one very
special category of points, therefore this restriction does not
have an impact on our results.\footnote{This category of points will be
  introduced later and can be seen in the bottom left plot of
 \reffig{fig:hew_s:notDZTP}.}

\begin{table}[t]
  \centering
  \begin{tabular}{ccc}
    \toprule
    Parameter & Range & Metric\\
    \midrule
    $\lambda$ & $0,\, \pi/2$ & flat\\
    $|\kappa|$ & $0,\, \pi/2$ & flat\\
    $|A_\lambda|$ & $0,\, 10\tev$ & hybrid\\
    $|A_\kappa|$ & $0,\, 10\tev$ & hybrid\\
    $|A_t|$ & $0,\, 10\tev$ & hybrid\\
    $\msusy$ & $1,\, 10\tev$ & log \\
    $|v_S|$ & $0,\, 10\tev$ & hybrid\\
    $\tan\beta$ & $1,\, 60$ & log\\
    \bottomrule
  \end{tabular}
  \caption{Ranges and metric of parameters that we scanned in the NMSSM at the SUSY scale. We considered positive and negative $\kappa$, $v_S$ and trilinear couplings. The ``hybrid'' metric is flat below 10\gev, and logarithmic elsewhere. The top mass was fixed to its measured value $173.1\gev$~\cite{Patrignani:2016xqp}.}
  \label{Tab:scan}
\end{table}

The main experimental constraints on the parameter region of interest come from LEP chargino searches and the observed Higgs properties.
The Higgs sector of our model must be compatible with observations of an SM-like Higgs boson with a mass close to $125\gev$. The observed Higgs, however, could correspond to any one of the three neutral Higgs bosons in our model. We calculated tree-level reduced couplings between the neutral Higgs bosons and SM fermions by taking into account mixing between the neutral Higgs bosons. We furthermore calculated one-loop reduced couplings between the Higgs bosons and photons and gluons using \FS routines developed in \refcite{Staub:2016dxq}. By passing this information and the Higgs masses to \texttt{Lilith-1.1.4\_DB-17.05}~\cite{Bernon:2015hsa}, we find a chi-squared, $\chi^2_\text{Higgs}$, for our Higgs sector from Run I and II measurements of the Higgs boson at the LHC.

We penalized points in tension with LEP bounds on charginos~\cite{Patrignani:2016xqp} by introducing a chi-squared for the effective $\mu$-parameter
\begin{equation}
\chi^2_\text{LEP} \equiv
\begin{cases}
0 & \mueff \ge 100\gev , \\
\left(\frac{\mueff - 100\gev}{5\gev}\right)^2  & \mueff < 100\gev .
\end{cases}
\end{equation}
We constructed this function to guide our sampling algorithm towards acceptable solutions with $m_{\tilde\chi^\pm_1} \gtrsim 100\gev$, rather than precisely reflect experimental constraints from LEP. We furthermore penalized points without an SFOPT by the chi-squared
\begin{equation}
\chi^2_\text{SFOPT} \equiv \left(\frac{\log_{10}\gammaEW}{0.2}\right)^2 .
\end{equation}
The role of this term is to focus our sampling algorithm on SFOPT with $\gammaEW \simeq 1$; it is in fact equivalent to a Gaussian penalty $\log_{10}\gammaEW = 0 \pm 0.2$.

Since the parameter space shown in \reftab{Tab:scan} is eight-dimensional we sampled points from our model using \texttt{MultiNest-3.10}~\cite{Feroz:2007kg, Feroz:2008xx, 2013arXiv1306.2144F} with a chi-squared
\begin{equation}
\chi^2 = \chi^2_\text{Higgs} + \chi^2_\text{SFOPT} + \chi^2_\text{LEP} .
\end{equation}
We saved and considered all points evaluated by \texttt{MultiNest}, i.e., we disabled the cuts ordinarily placed on saved points by the \texttt{MultiNest} algorithm.
To be consistent with the LHC Higgs measurements and LEP bounds on charginos~\cite{Patrignani:2016xqp}, and to satisfy our SFOPT requirement, we select points with
\begin{equation}
\chi^2_\text{Higgs} - \min \chi^2_\text{Higgs}\le 6.18,~ \mueff \ge 100\gev ~\text{and}~ \gammaEW \ge 1 ,
\end{equation}
where $\min \chi^2_\text{Higgs}=22.3$ was the minimum $\chi^2_\text{Higgs}$ found in our scan. After that, we further required that remaining points satisfied LHC and LEP bounds on BSM Higgs bosons using \texttt{HiggsBounds-5.3.2beta}~\cite{arXiv:0811.4169,arXiv:1102.1898,arXiv:1301.2345,arXiv:1311.0055,arXiv:1507.06706}, which we interfaced via \texttt{NMSSMCALC}~\cite{Baglio:2013iia}.

\subsection{Classification of phase transitions}

After collecting more than three million valid points, we found that
the possible phase structures in the NMSSM
harbored rich and novel phenomenology.  To reflect this phenomenology,
we classify these points into three categories that differ by the
nature of the first possible PT in the cosmological
history:
\begin{enumerate}
  \item \TypeEWS: EW symmetry is spontaneously broken such that at least one Higgs field and the singlet field obtain non-vanishing VEVs simultaneously.
  \item \TypeEW: EW symmetry is spontaneously broken by one or both Higgs fields obtaining VEVs, but the singlet VEV remains zero.
  \item \TypeS: EW symmetry remains unbroken, but the singlet field obtains a VEV. The Higgs obtain non-vanishing VEVs in a SFOPT afterwards, during which the sign of singlet VEV may be maintained or flipped. Thus we further classify this type into two subcategories:
  \begin{itemize}
    \item \TypeSs: the strongest PT maintains the sign of singlet VEV.
    \item \TypeSf: the strongest PT flips the sign of singlet VEV.
  \end{itemize}
\end{enumerate}
It is important to understand that at this stage we do not have the means to ensure that a PT is definitely part of the cosmological history.  More precisely, for such an extensive sample of parameter points, we are not in the position to calculate nucleation temperatures, actions, decay rates, etc.\ for each potential transition
in the phase structure.  For this reason, unless specified otherwise when we say `PT' we typically mean `\textit{possible} PT'.

To simplify our discussion of this non-trivial structure, we introduce the following shorthand notation:
\begin{itemize}
  \item We denote the minimum value of the potential in a given direction with a calligraphic font.  For example, $\minima{s}$ is a value of singlet field $s$ at a minimum of the scalar potential.
  \item By the triplet of values e.g., $(100, 200, 300)$, we mean $\minima{h}_u=100\gev$, $\minima{h}_d=200\gev$, and $\minima{s}=300\gev$.
  \item At a critical temperature, two vacua are degenerate. However, we always define the true vacuum to be the deepest of these vacua just below the critical temperature, and the other one is the false vacuum in our notation.
  \item In case of multiple SFOPTs we refer to the SFOPT with the greatest \gammaEW as the strongest one.
  \item We define
    \begin{equation}
    \minima{h} \equiv \sign(\minima{h}_u\minima{h}_d) \sqrt{\minima{h}_u^2+\minima{h}_d^2} .
    \end{equation}
    as the signed geometric mean of the Higgs fields.
\end{itemize}

\subsection{Benchmark points}

\begin{figure}[ht]
\centering
\includegraphics[width=.9\textwidth]{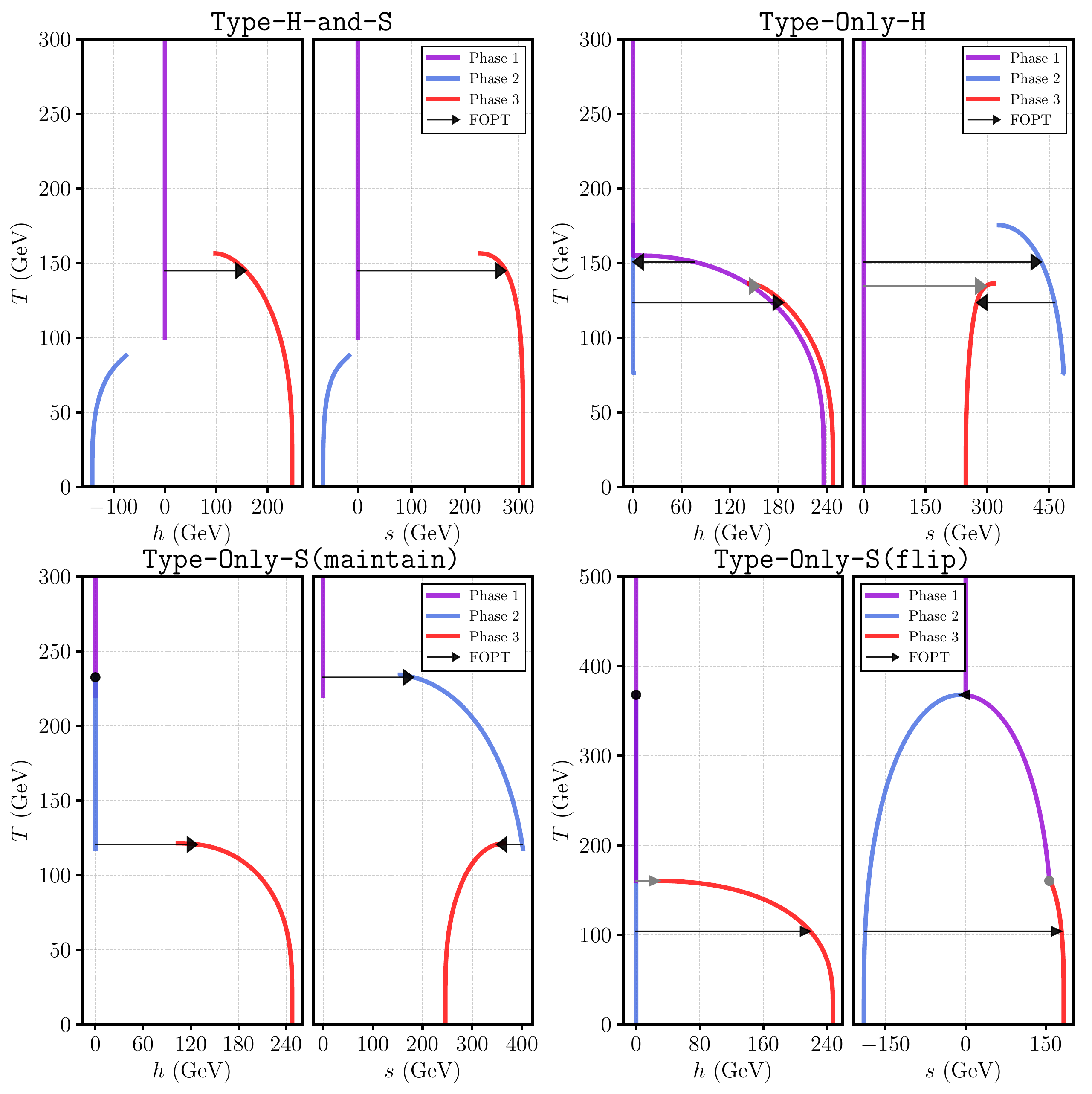}
\caption{Phase structures for typical points in the categories \TypeEWS (upper left), \TypeEW (upper right), \TypeSs (lower left) and \TypeSf (lower right). The lines show the field values at a particular minimum as a function of temperature. The arrows indicate that at that temperature the two phases linked by the arrows are degenerate and thus that a FOPT could occur in the direction of the arrow.  The dots in the lower panels represent transitions that do not change the corresponding field values. The black arrows and dots show a path that includes the strongest EW FOPT, while the gray ones are not in that path.}
\label{fig:benchmarks_1}
\end{figure}

In \reffig{fig:benchmarks_1}, we present a phase history for a typical point in each category. For these benchmark points, we checked our results with \cosmo and calculated the nucleation temperature for every possible FOPT. The corresponding input parameters, Higgs properties and transition information are shown in \reftab{table:bk}. On each panel, the lines show the signed geometric mean of the Higgs fields (left) or the singlet field (right) at a minimum of the potential as a function of temperature.\footnote{Note though that two phases connected by crossover PTs are merged into one phase in order to simplify the phase structure.} 
Two phases linked by an arrow at a given temperature are degenerate and thus a FOPT could occur in the direction indicated by the arrow (i.e., below the critical temperature, the phase at the end of the arrow contains a deeper minimum). 
When there is more than one possible sequence of FOPTs that leads from the origin at $T=1\tev$ to the observed vacuum at $T=0$, we show the FOPTs that belong to the sequence that includes the strongest FOPT by black arrows, and PTs that are not part of that history by gray arrows. 
Note though that other possible FOPTs between phases that are never degenerate are not marked. For example, in the upper left panel, the minima in phase 2, which appears at about $T=88\gev$, always lies shallower than that in phase 3. A FOPT between them is possible, although there is no critical temperature.

{
\newcommand{\mc}[3]{\multicolumn{#1}{#2}{#3}}
\newcommand{\mr}[3]{\multirow{#1}{#2}{#3}}
\newcommand{\ssk}{\shortstack}
\newcommand{\ce}{\centering}
\renewcommand{\globalmin}{SM vac.}
\begin{table}[ht]
\centering
\begin{tabular}{lcccc}
\toprule
                                 & \TypeEWS 								& \TypeEW   				& \mc{1}{p{2.5cm}}{\ce \TypeSs}	& \mc{1}{p{2.5cm}}{\ce \TypeSf}\\
\midrule								
$\lambda$                        	&  0.618   								&  0.607   				&    0.601    							& 0.935     \\
$\kappa$                         	&  0.229   								&  0.191   				&    0.175    							& 1.137     \\
$A_\lambda$                      	&  160.1   								&  160.5   				&  170.0    								& 147.4     \\
$A_\kappa$                       	&  $-93.7$ 								&  $-117.5$				& $-25.2$  								& 61.4      \\
$A_t$                            	&  $-21.4$ 								&  38.3	   				& $-24.6$  								& $-478.6$  \\
$\msusy$                         	&  6374.7  								&  3463.1 				& 5857.5   								& 4164.3    \\
$v_S$                            	&  307.9   								&  247.5			  		&  245.7    								& 183.1     \\
$\tan\beta$                      	&  1.2     								&  2.0						&    2.6      							& 3.2       \\
\midrule								
$m_{H_1}$                        	& 91.7     								&  47.9					&   45.6    								& 126.2   \\
$m_{H_2}$                        	& 127.4    								&  124.6					&  125.1    								& 184.4   \\
$m_{H_3}$                        	& 237.6    								&  226.6					&  252.7    								& 366.5   \\
$m_{A_1}$                        	& 167.3    								&  145.9					&  103.8    								& 145.4   \\
$m_{A_2}$                        	& 229.7    								&  225.9					&  248.2    								& 325.8   \\
$m_{H^{\pm}}$                    	& 214.2    								&  206.7					&  233.1    								& 294.3   \\
$\chi^2_\text{Higgs}$            	& 27.0     								&  25.6					&   26.2    								& 26.4    \\
\midrule
\multicolumn{5}{c}{First PT}\\
\midrule
Order                            	& 1st 										& 2nd at $T=155$		& 1st 										& 1st \\
False vac.                       	& $(0,0,0)$								&    $(0,0,0)$			& $(0,0,0)$								& $(0,0,0)$   \\
True vac.                      	& $(106, 117, 276)$					& $(0,\text{+ve},0)$	& $(0,0, 182)$							&  $(0,0,-12)$ \\
$T_C$                         	   & 145      								&     N/A          	& 233        							&  368       \\
$T_N$				                 	& 116        							&     N/A 				& 230	 									&	367	\\
\midrule
\multicolumn{5}{c}{Strongest FOPT}\\
\midrule
False vac.                       	&\mr{4}{*}{\ssk{Same\\as above}}	& $(0, 0,463)$      	& $(0,0,400)$ 							& $(0,0,-188)$   \\
True vac.                    		&                        			& $(91, 162, 274)$  	&$(59,114,349)$							& $(66,209,179)$ \\
$T_C$                            	&       									& 124          			& 121          							& 104          \\
$T_N$                            	&                        			& 119            		& 119          							& N/A; no nuc. \\
$\gammaEW$                       	&  1.1                  				& 1.5              	&   1.1      							& 2.1            \\
Ends at \globalmin 					& Yes 										& Yes 						& Yes 										& Yes  \\
Possible 									& Yes 										& No; prior PT fails	& Yes 										& No; no nuc.\\
\bottomrule
\end{tabular}
\caption{Benchmark points for our four scenarios. All dimensionful quantities are in GeV.
The abbreviation vac.\ is for vacuum and nuc.\ is for nucleation.
The \text{+ve} in \TypeEW means that the field value of vacuum during the 2nd order phase transition is shifted to positive direction.}
\label{table:bk}
\end{table}
}

From \reffig{fig:benchmarks_1} we can see that in all categories at high temperature, $T>400\gev$, the true vacuum is always at the origin (as described in \refsec{sec:FOPT} this is a requirement in our scan). In the upper left panel, the first (and only) PT occurs at $T \lesssim 145\gev$ between $(0,0,0)$ and $(106, 117, 276)$ with $\gammaEW=1.09$ and nucleation temperature $T_N=116\gev$. Thus it is classified as \TypeEWS.

In the upper right panel, only one of the Higgs fields, $h_u$, develops a VEV during the first crossover transition at $T = 155\gev$. The first transition in the cosmological history was never first order in our \TypeEW samples. As the Universe cools, however, a deeper minimum exists between $T=151\gev$ and $T=124\gev$ at about $(0, 0, 450)$, which belongs to phase 2. The FOPT to this deeper minimum would (temporarily) restore EW symmetry; however, we find that it cannot complete as~\eqref{eq:define_TN} cannot be satisfied.
If it completed, EW symmetry would subsequently be permanently broken by another SFOPT at $T \lesssim 123.6\gev$ which would complete, from $(0,0,463)$ to $(91,162,274)$ with $\gammaEW=1.5$ and $T_N=119\gev$. 
Indeed, in all the \TypeEW samples that we found, EW symmetry was broken, possibly restored and finally broken again, and the {\it final} FOPT would be the strongest, just as in this example. However, these sequences of transitions are impossible, as the actions for the transitions that restore EW symmetry are always so large that bubbles cannot nucleate properly.
Thus although there appear to be SFOPTs with $\gammaEW>1$ and nucleation temperatures in the \TypeEW samples, they cannot explain the observed baryon asymmetry of the Universe, as a previous transition in the cosmological history would not complete.

For the \TypeSs point (lower left panel) in the first transition at $T=233\gev$ only the singlet obtains a positive VEV; EW symmetry is broken with the sign of singlet VEV maintained in the second (and final) PT at $T=121\gev$. Both of the transitions are strongly first order and complete.  Although transitions in which only the singlet obtains a VEV cannot precipitate baryogenesis, they might nevertheless result in interesting gravitational wave signatures.

Finally, we consider a \TypeSf point (lower right panel). The singlet field develops a negative value during the first transition at $T=368\gev$, which is first-order and completes at $T_N = 367\gev$. At $T\lesssim 368\gev$, just below the critical temperature of the first transition, a phase with positive $\minima{s}$ develops, which is approximately symmetric with respect to the phase with negative $\minima{s}$. Eventually, a second first-order transition at $T=104\gev$ breaks EW symmetry and flips the sign of the singlet by transitioning to this approximately symmetric phase. Although this is the strongest PT, it cannot complete, as the barrier between the phases means that the tunneling action is too large for~\eqref{eq:define_TN} to be satisfied. This phenomenon appears in a large fraction of our \TypeSf samples.

Phase histories of types \TypeEWS and \TypeSs were previously investigated in \refcite{Balazs:2013cia,Cheung:2012pg,Huang:2014ifa,Kozaczuk:2014kva}; however, the richer phase histories in \TypeEW and \TypeSf have not been discussed in the literature as far as we are aware. Note that the barrier between the minima in the \TypeEW and \TypeSf are usually so high that the tunneling may not happen. This shows the importance of studying phase structure as well as calculating the transition strength.

We also checked the robustness of our results against the change of
the renormalization scale. For the \TypeEWS benchmark point in
\reftab{table:bk}, we found a mild (1\% -- 2\%) variation of the
critical temperature and the transition strength as the
renormalization scale changes in the $(m_t/2,2m_t)$ range. We
furthermore checked gauge dependence by repeating our calculations for
our benchmark points in the $\xi = 0$ (Landau) gauge. We found, as
anticipated, that gauge dependence was present but typically mild,
especially for the critical temperatures.  The gauge dependence could,
nevertheless, motivate the application of gauge independent techniques
in future works.

\subsection{Reaching the observed \globalmin}

During the scan we required that the deepest minimum at zero temperature agreed with the observed VEV, $\minima{h} = 246\gev$. We call the phase associated with the observed VEV the \globalmin. We split our samples by two ways of reaching the \globalmin. First, in \refsec{sec:simply_case} we consider samples for which the strongest SFOPT ends in the \globalmin, which changes smoothly to $\minima{h} = 246\gev$ at $T=0$. Second, in \refsec{sec:complex_case} we consider samples for which the strongest FOPT does not end in the \globalmin. As we discuss, such samples must feature at least one further FOPT that ultimately ends in the observed vacuum at $T=0$. In both cases, the \TypeEW scenario was by far the rarest, with noticeably few samples shown in the following scatter plots.

\subsubsection{The strongest FOPT ends in the \globalmin}\label{sec:simply_case}

\begin{figure}[t]
\centering
\includegraphics[width=.9\textwidth]{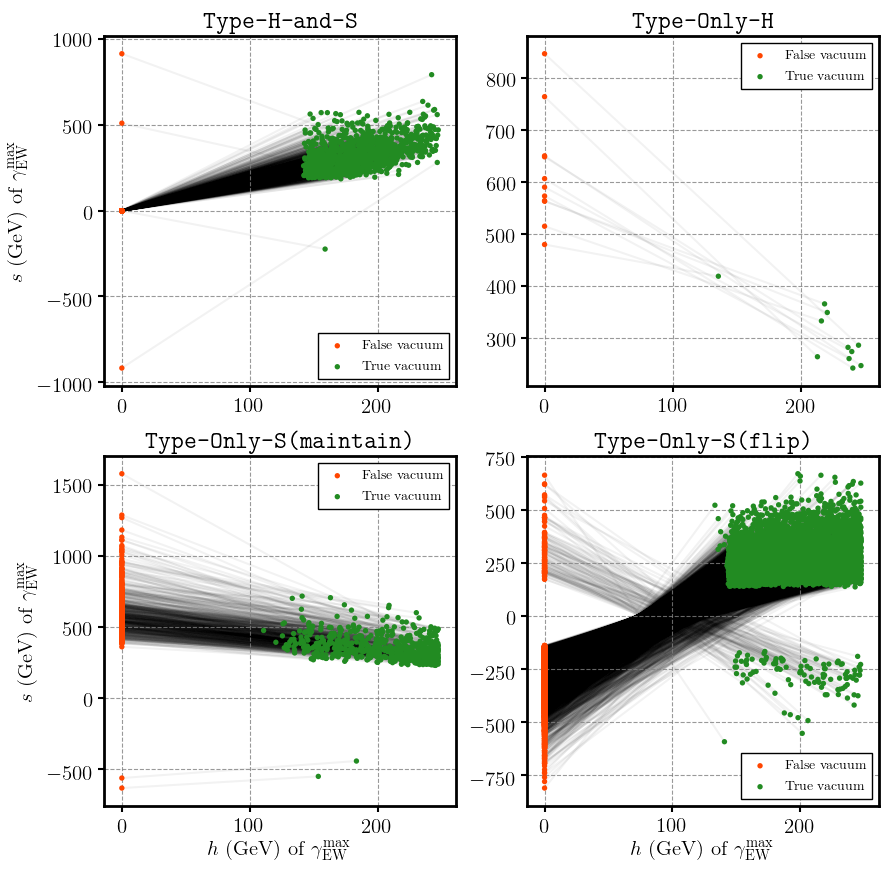}
\caption{The Higgs and singlet field values at the true and false minima at the critical temperature of the strongest FOPT for samples for which the strongest FOPT ends in the \globalmin.}
\label{fig:hew_s:DZTP}
\end{figure}

We selected samples in which the strongest FOPT ended in the \globalmin. For our samples, it was sufficient to check that $\minima{h}_u>0\gev$ and $\minima{h}_d>0\gev$ for the true vacuum of strongest FOPT. All of our benchmark points in \reftab{table:bk} are in this category.
In \reffig{fig:hew_s:DZTP}, we present the true and false minima of the {\it strongest} FOPT at the critical temperature. It demonstrates some features of each of the types of point that we described above. For \TypeEWS, the {\it first} transition, in which the Higgs and singlet fields acquire VEVs, is usually also the {\it strongest} FOPT. There are however three exceptional points where the singlet field values at the false minimum are non-zero. They have similar phase structures to the upper right panel of \reffig{fig:benchmarks_1} except that the minima of phase 1 is always deeper than the minima of phase 2 in all three cases. Thus there is no critical temperature between these phases, and so the strongest FOPT for these three points is not in the cosmological history.

According to the definition of \TypeEW, only the Higgs fields obtain VEVs in the first transition in the history, while the upper right panel of \reffig{fig:hew_s:DZTP} shows that the false vacuums of strongest FOPT have zero Higgs VEVs, $\minima{h}_u=\minima{h}_d=0$, but a non-vanishing singlet VEV, $\minima{s}\neq 0$. This means that there must be an intermediate transition that restores EW symmetry and generates a singlet VEV. Since the number of \TypeEW scenarios  that we found are quite small, we checked each one in detail.  We found that this intermediate transition exists for all \TypeEW samples, but the corresponding tunneling probabilities are too small.  Nonetheless it is possible that there exist scenarios of this type where the transition does complete. 

The lower panels of \reffig{fig:hew_s:DZTP} display samples of \TypeS where the strongest FOPT maintains (left) or flips (right) the sign of the singlet VEV. We see that the singlet VEV can evolve to up to $1.6\tev$ after the first transition, and then shifts to about $150\gev$ to $650\gev$ during the strongest FOPT. The singlet VEV $\minima{s}$ of the true vacuum can be both positive or negative, because the input $v_S$ includes both signs. We have checked that the singlet VEV at the true vacuum has the same sign as the input $v_S$.

\begin{figure}[t]
\centering
\includegraphics[width=.9\textwidth]{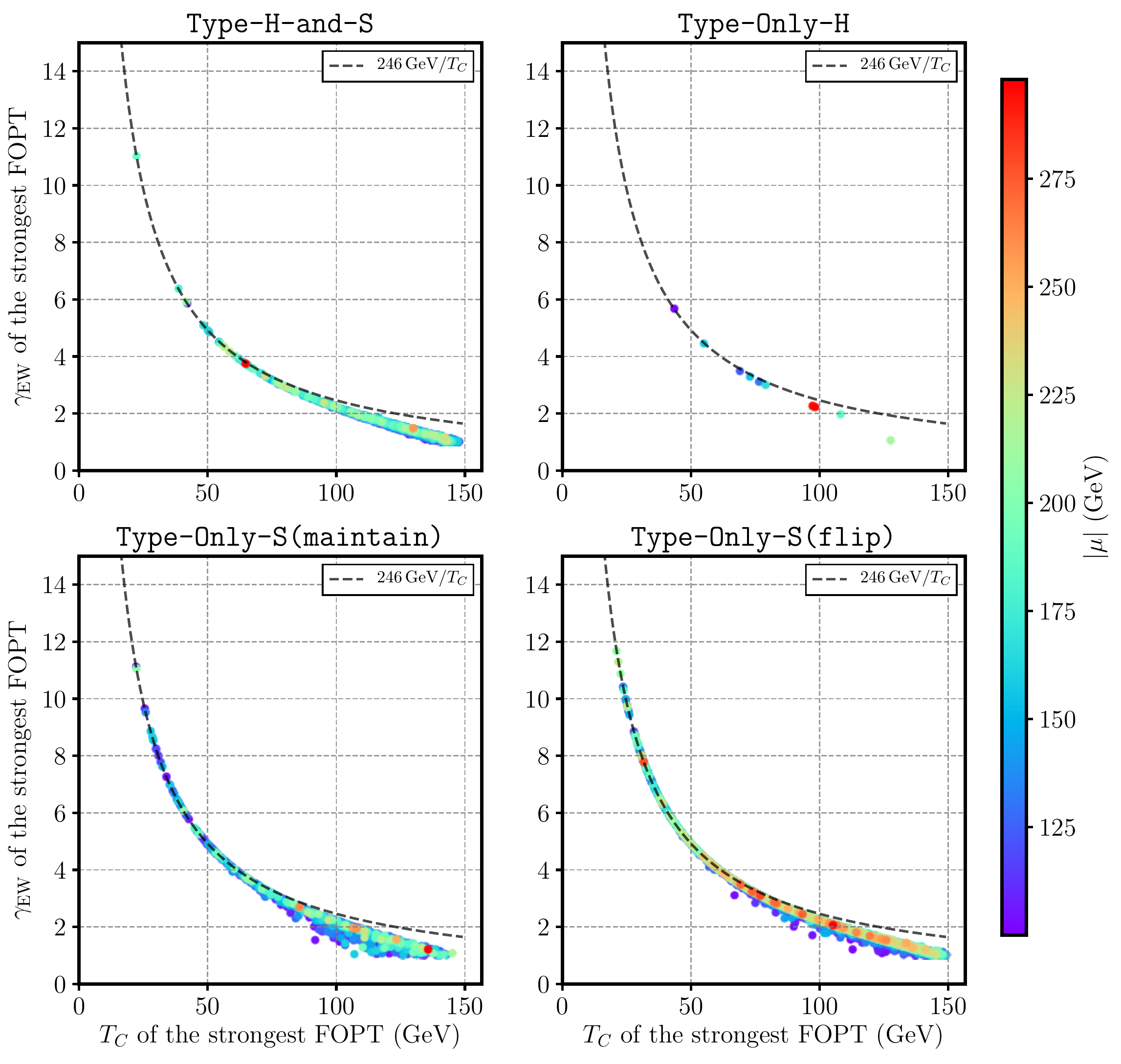}
\caption{The critical temperature and order parameter for the strongest PTs for samples for which the strongest FOPT ends in the \globalmin. The points are colored by the effective $\mu$-parameter.}
\label{fig:Tc_gamma:DZTP}
\end{figure}

In all scenarios, the spread in the possible true vacuum for the Higgs fields at the critical temperature is small, and typically it matches and rarely exceeds the input EWSB vacuum, i.e., $\minima{h} \lesssim 246\gev$. This can be further seen in \reffig{fig:Tc_gamma:DZTP}, which shows the FOPT strength against the critical temperature. The strength lies close to what it would be if $\minima{h} = 246\gev$ (dashed gray line). For higher critical temperatures, however, deviations from the gray line are visible, as the thermal loop-corrections are relevant. The thermal loop-corrections tend to make the potential more convex, thus decreasing $\minima{h}$ at the critical temperature and the strength of the PT.

\begin{figure}[t]
\centering
{\includegraphics[width=.9\textwidth]{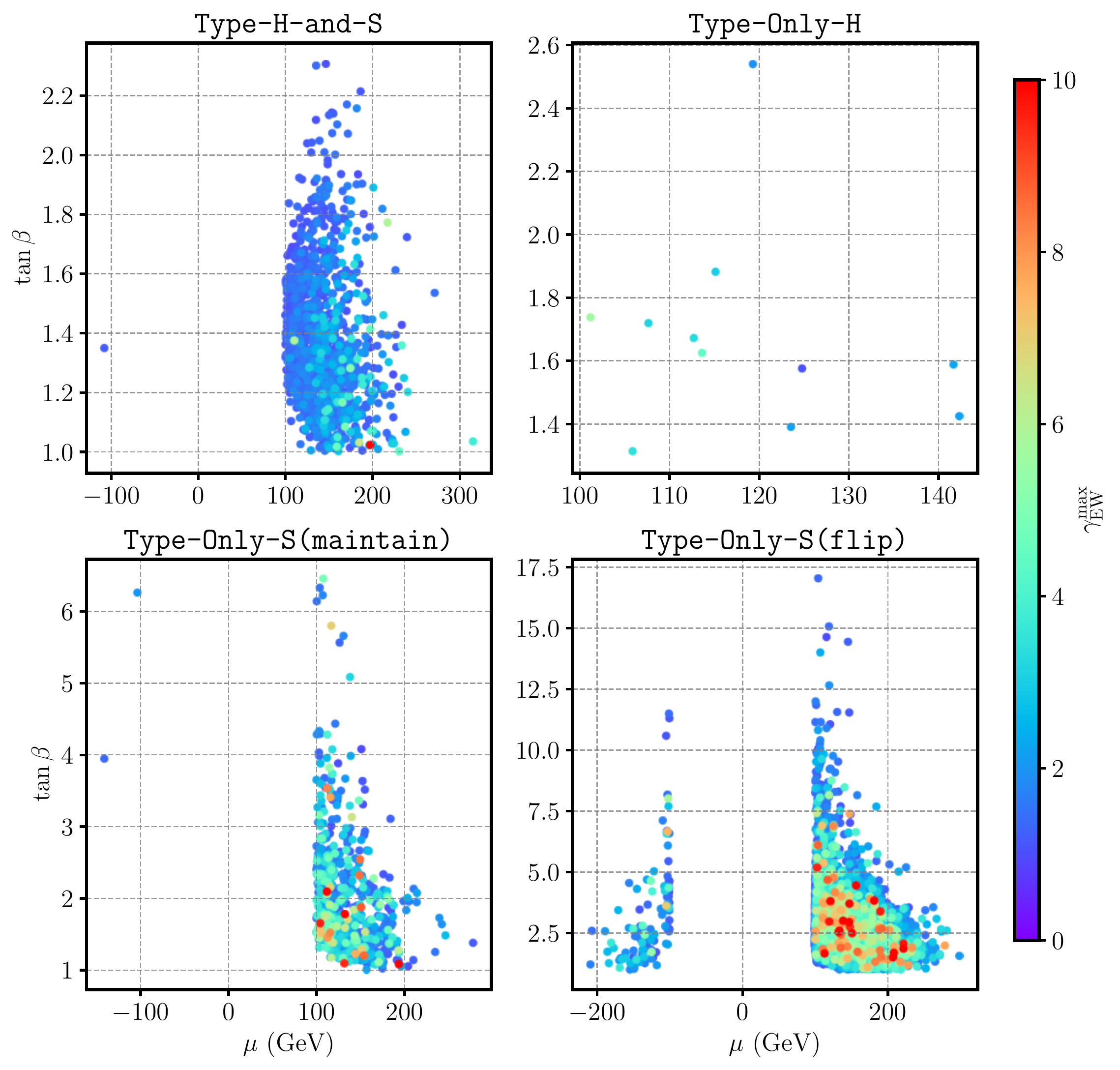}}
\caption{The parameters $(\mueff,\tan\beta)$ for samples for which the strongest FOPT ends in the \globalmin. The points are colored by the $\gammaEW$ of the strongest FOPT.}
\label{fig:mu_tanb:DZTP}
\end{figure}

We now delineate the regions of the NMSSM parameter space in which our four scenarios occur.
We checked that in all scenarios the stops were truly decoupled by checking stop mixing, $X_t = A_t - \mueff \cot\beta$, which could potentially split the stop mass eigenvalues making one of them light. We found that most samples were actually concentrated within the range $-\msusy \leq X_t \leq \msusy$ and no particular value of $\msusy$ was preferred by our samples. 

In \reffig{fig:mu_tanb:DZTP} we show that the Higgs sector parameters ($\mueff$, $\tan\beta$) are severely constrained. Indeed, the \TypeEWS and \TypeEW scenarios require $\tan\beta\lesssim3$, whereas the \TypeSs and \TypeSf scenarios permit $\tan\beta\lesssim7$ and $\tan\beta\lesssim17$, respectively. For all types, the upper limit of $\tan\beta$ decreases with $\mueff$ increasing. The effective $\mu$-parameter, and thus the higgsinos,  are always light, $|\mueff| \lesssim 300\gev$. Thus we find further motivation for scenarios with small $\mueff \lesssim 1\tev$, which are also motivated by naturalness, and we anticipate that the searches for higgsinos at the LHC could be sensitive to our models. Samples with $\mueff < 0$ were extremely rare in the \TypeEWS and \TypeSs scenarios, and not present in the \TypeEW samples. 

\begin{figure}[t]
\centering
{\includegraphics[width=.9\textwidth]{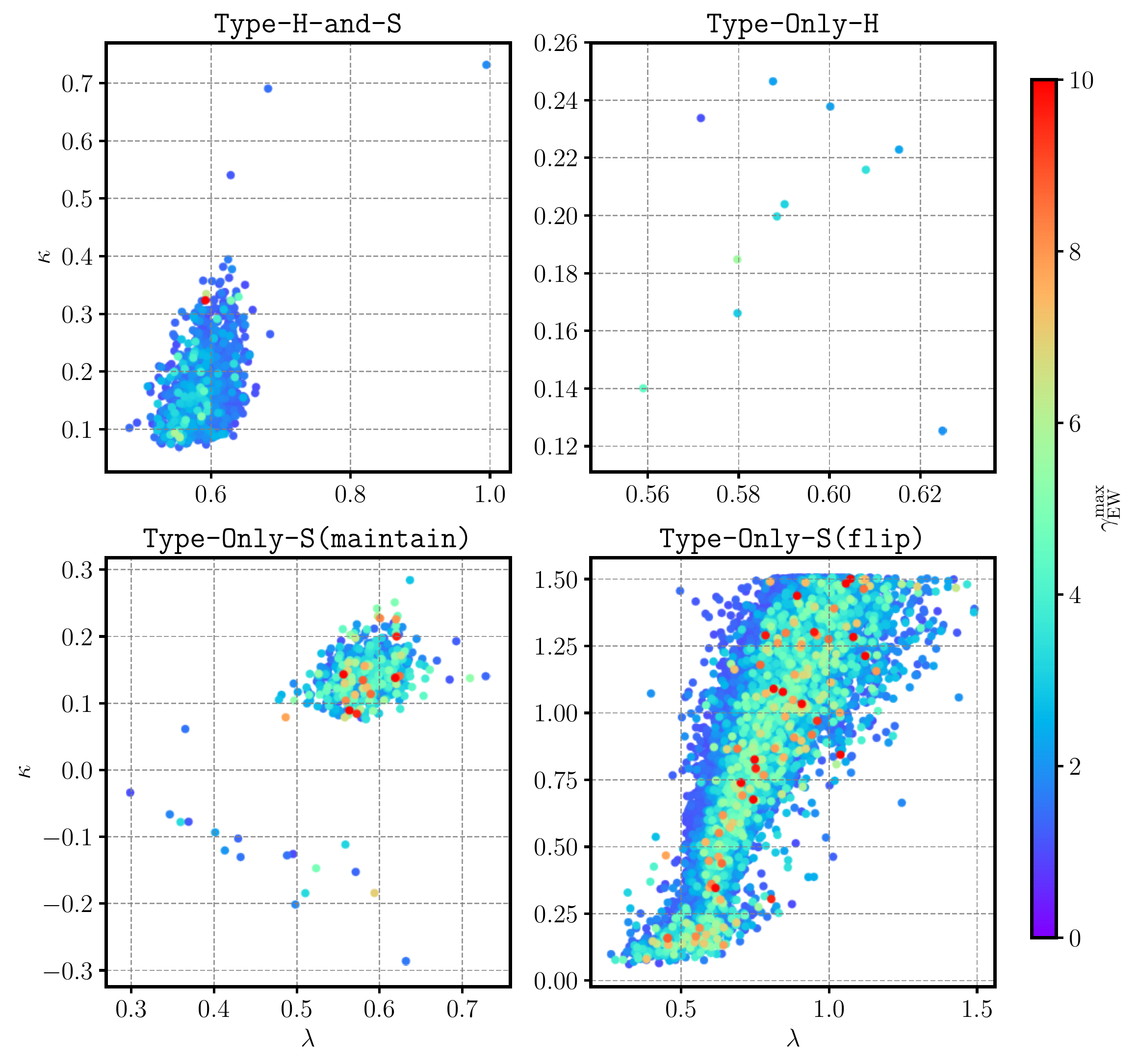}}
\caption{The quartics $(\lambda,\kappa)$ for samples for which the strongest FOPT ends in the \globalmin. The points are colored by the $\gammaEW$ of the strongest FOPT.}
\label{fig:lambda_kappa:DZTP}
\end{figure}

We see, furthermore, in \reffig{fig:lambda_kappa:DZTP}, that quartic couplings of around $\lambda \approx 0.6$ and $\kappa \approx 0.2$ could result in an SFOPT in all our scenarios, though a broad range of couplings result in SFOPTs in \TypeSf scenario, including couplings with values far above the limits that would be set if we required perturbativity up to the GUT scale. The constraints strongly prefer that $\lambda \kappa > 0$, a combination that is invariant under the field redefinition $S \to -S$. Since we worked in a $\lambda > 0$ convention, the inequality $\lambda \kappa > 0$ is equivalent to $\kappa > 0$. In the \TypeSs scenarios, however, we find a few solutions for which  $\kappa < 0$. 

\begin{figure}[t]
\centering
{\includegraphics[width=.9\textwidth]{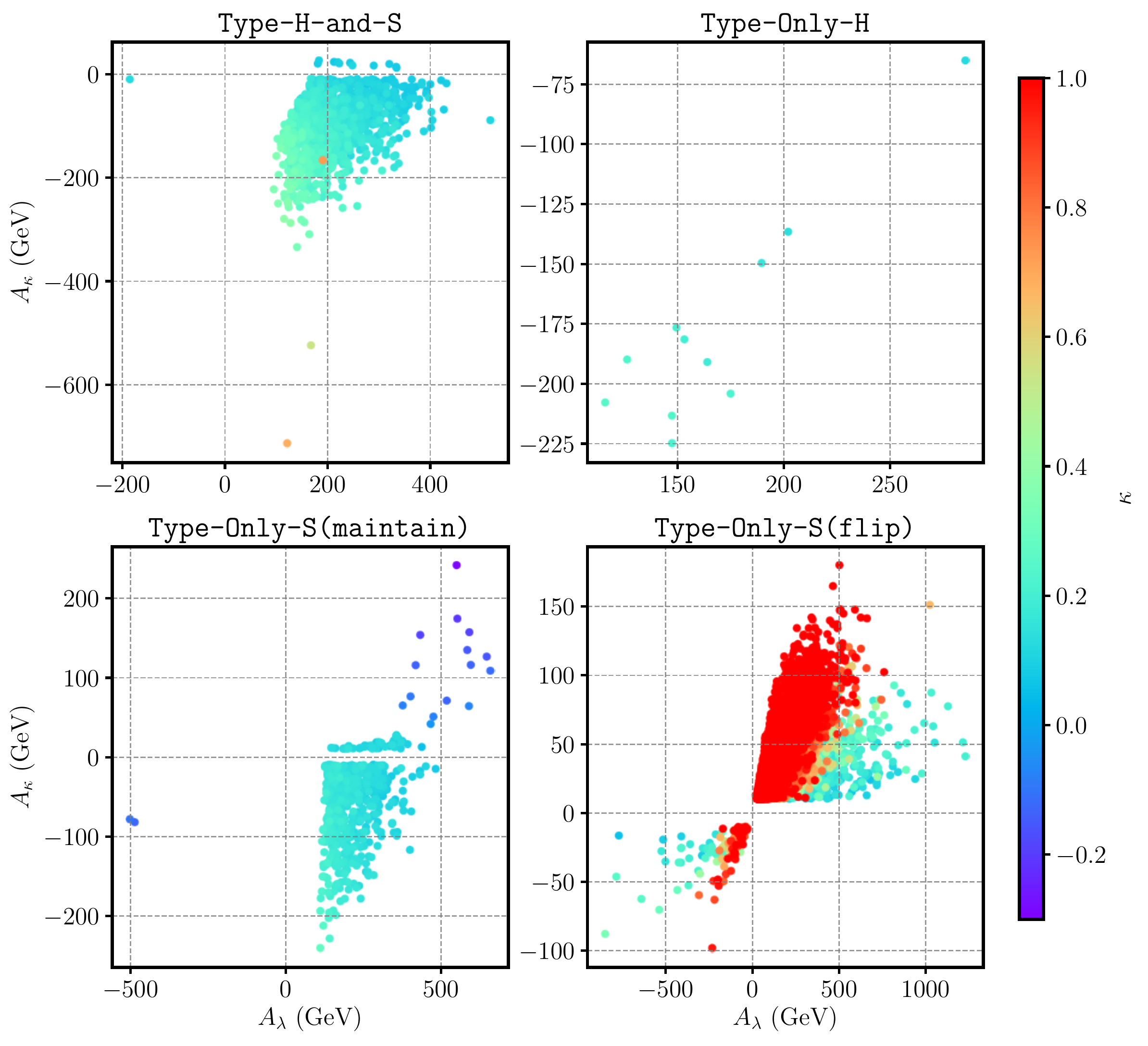}}
\caption{The trilinears $(A_\lambda, A_\kappa)$ for samples for which the strongest FOPT ends in the \globalmin. The points are colored by the parameter $\kappa$.}
\label{fig:Al_Ak:DZTP}
\end{figure}

\reffig{fig:Al_Ak:DZTP} shows the trilinear couplings ($A_\lambda$, $A_\kappa$) with the quartic coupling $\kappa$ shown by the color bar. The trilinears play an important role. As different types of sample require different sign of singlet VEV at low temperatures, the parameter space of each type shows distinguishable tendency. The samples in \TypeEWS, \TypeEW and \TypeSs scenarios are concentrated at negative $A_\kappa$ with positive $\kappa$ or positive $A_\kappa$ with negative $\kappa$, as well as a horizontal slice of points at $A_\kappa \approx 10\gev$ for \TypeEWS and \TypeSs. On the other hand, $A_\lambda$ is typically positive but $\lesssim 500\gev$. The one point with negative $A_\lambda$ in \TypeEWS and the two points with negative $A_\lambda$ in \TypeSs correspond the point of negative $\mueff$ in \reffig{fig:mu_tanb:DZTP}. The distinction between \TypeEWS and \TypeSs is that \TypeSs favors smaller $A_\kappa$ and $A_\lambda$. Finally, \TypeSf shows two approximately symmetric regions that were previously identified in \reffig{fig:mu_tanb:DZTP} by the sign of $\mueff$. The region of positive (negative) $A_\lambda$ and $A_\kappa$ corresponds to positive (negative) $\mueff$. 

We emphasize again that the parameter spaces shown in \reffig{fig:mu_tanb:DZTP}, \reffig{fig:lambda_kappa:DZTP} and \reffig{fig:Al_Ak:DZTP} can only ensure the existence of a SFOPT with $\gammaEW \gtrsim 1$. Establishing whether this SFOPT is definitely part of the cosmological history requires further investigation, which we only present for our benchmark points.

\subsubsection{The strongest FOPT does not end in the \globalmin}\label{sec:complex_case}

\begin{figure}[t]
\centering
\includegraphics[width=.9\textwidth]{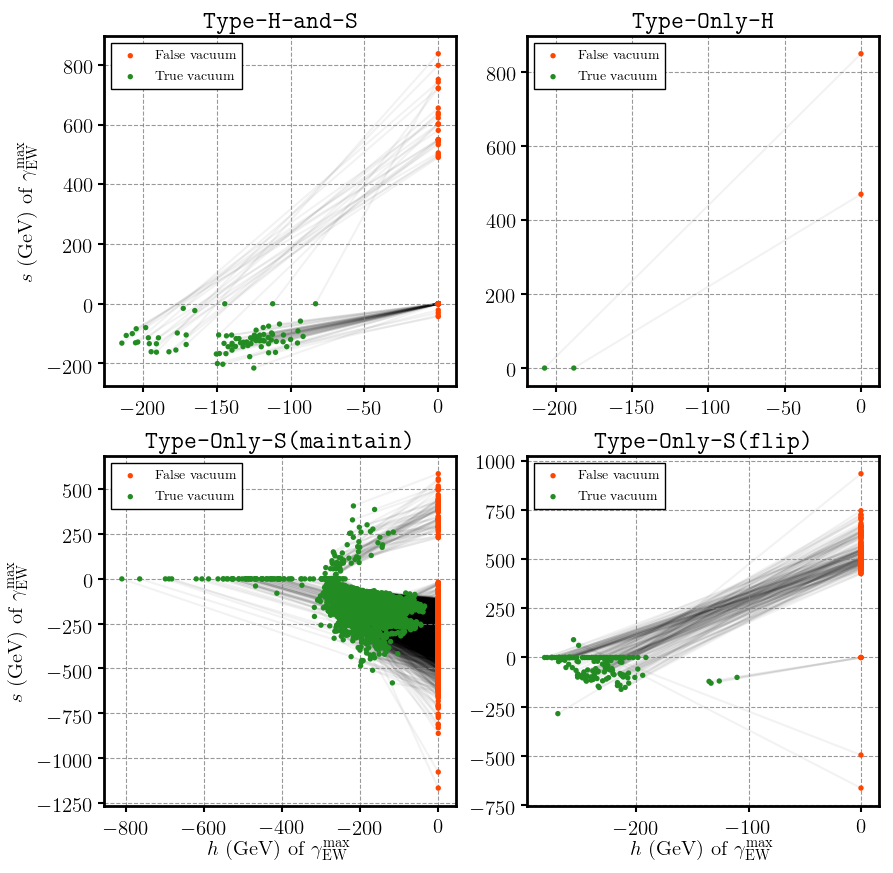}
\caption{The Higgs and singlet fields at the critical temperature of the strongest FOPT for samples for which the strongest FOPT does not end in the \globalmin.}
\label{fig:hew_s:notDZTP}
\end{figure}

\begin{figure}[t]
\centering
\includegraphics[width=.9\textwidth]{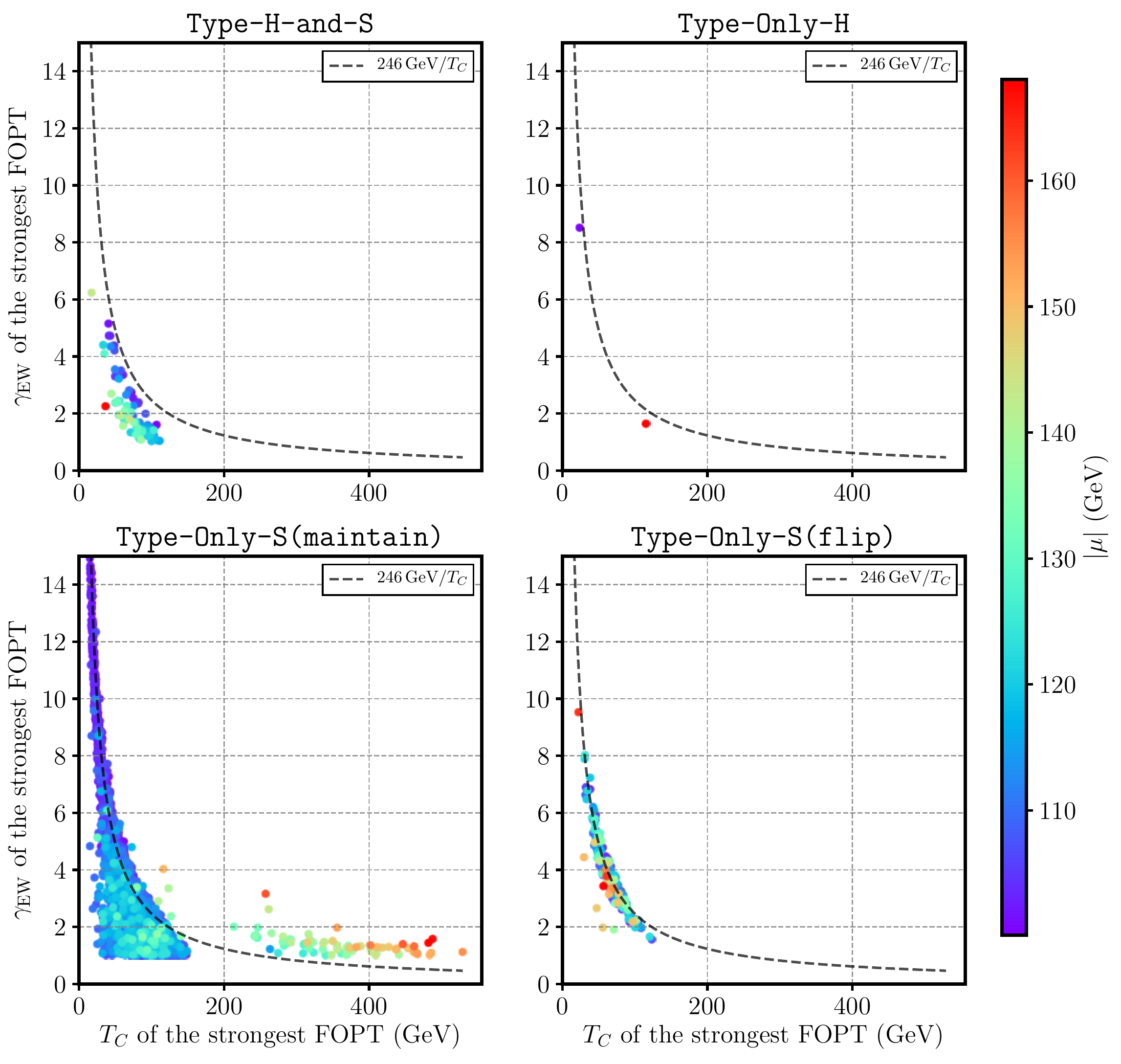}
\caption{The critical temperature and order parameter for the strongest PTs for samples for which the strongest FOPT does not end in the \globalmin. The points are colored by the effective $\mu$-parameter.}
\label{fig:Tc_gamma:notDZTP}
\end{figure}

Other than the scenario discussed above, we have plenty of samples in which the strongest FOPT does not end in the \globalmin, as shown in \reffig{fig:hew_s:notDZTP}. In these samples, in the true vacuum for the strongest FOPT, $\minima{h}$ is always negative and $\minima{s}$ is either zero or has a different sign to $\mueff$, so this almost certainly does not belong to the \globalmin in which $\minima{h} = 246\gev$.
The spread in the possible true vacuum for the Higgs fields at the critical temperature is substantial, and could differ considerably from the observed EW vacuum. Because of this, we no longer find that $\minima{h} \approx 246\gev$, allowing enhancement or suppression of the strength of the PT in \reffig{fig:Tc_gamma:notDZTP}, which differs markedly from \reffig{fig:Tc_gamma:DZTP}. Indeed, in the \TypeSs scenario, SFOPTs are possible for substantial critical temperatures of up to $T_C \lesssim 500\gev$.

\begin{figure}[t]
\centering
\includegraphics[height=.35\textwidth]{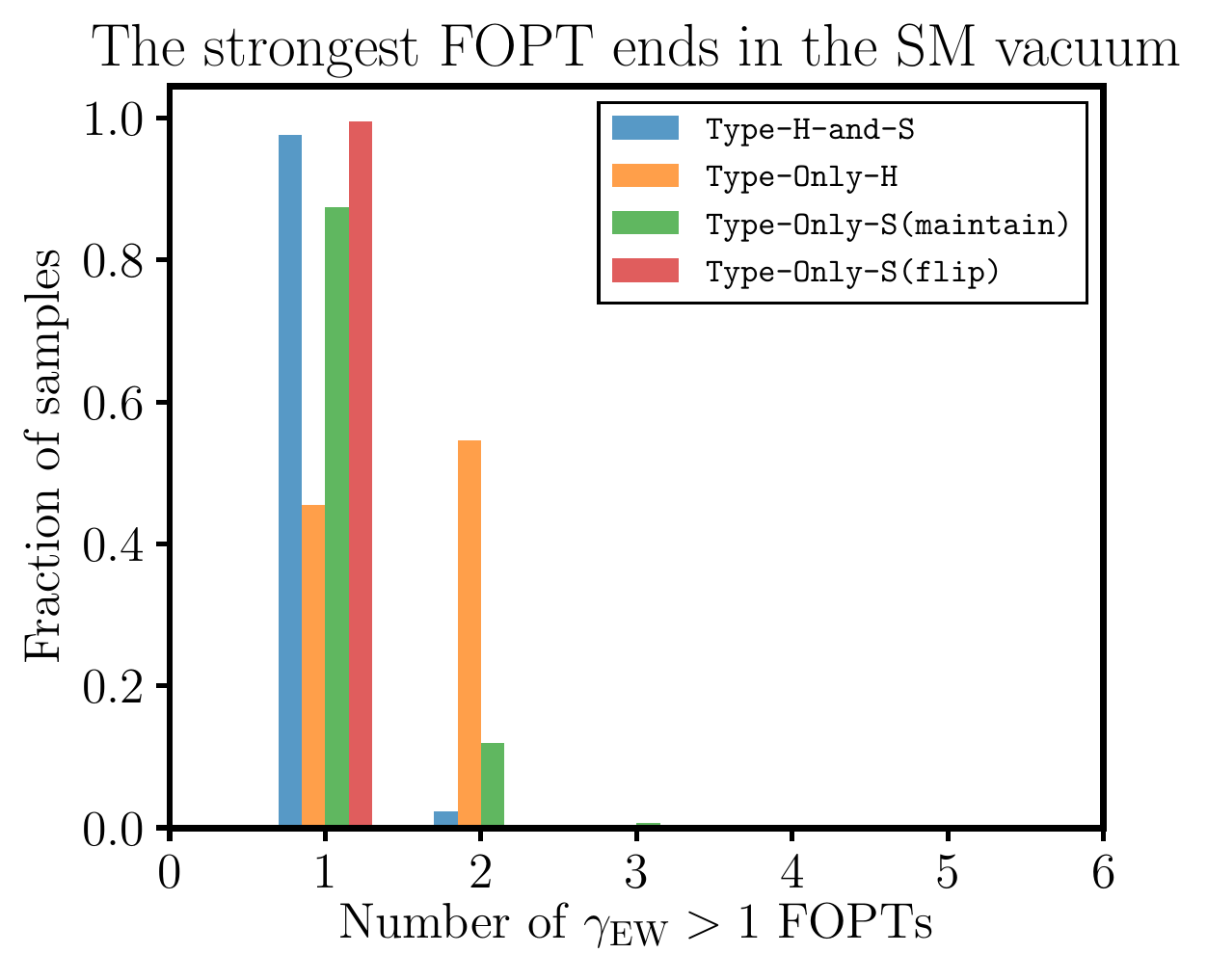}
\includegraphics[height=.35\textwidth]{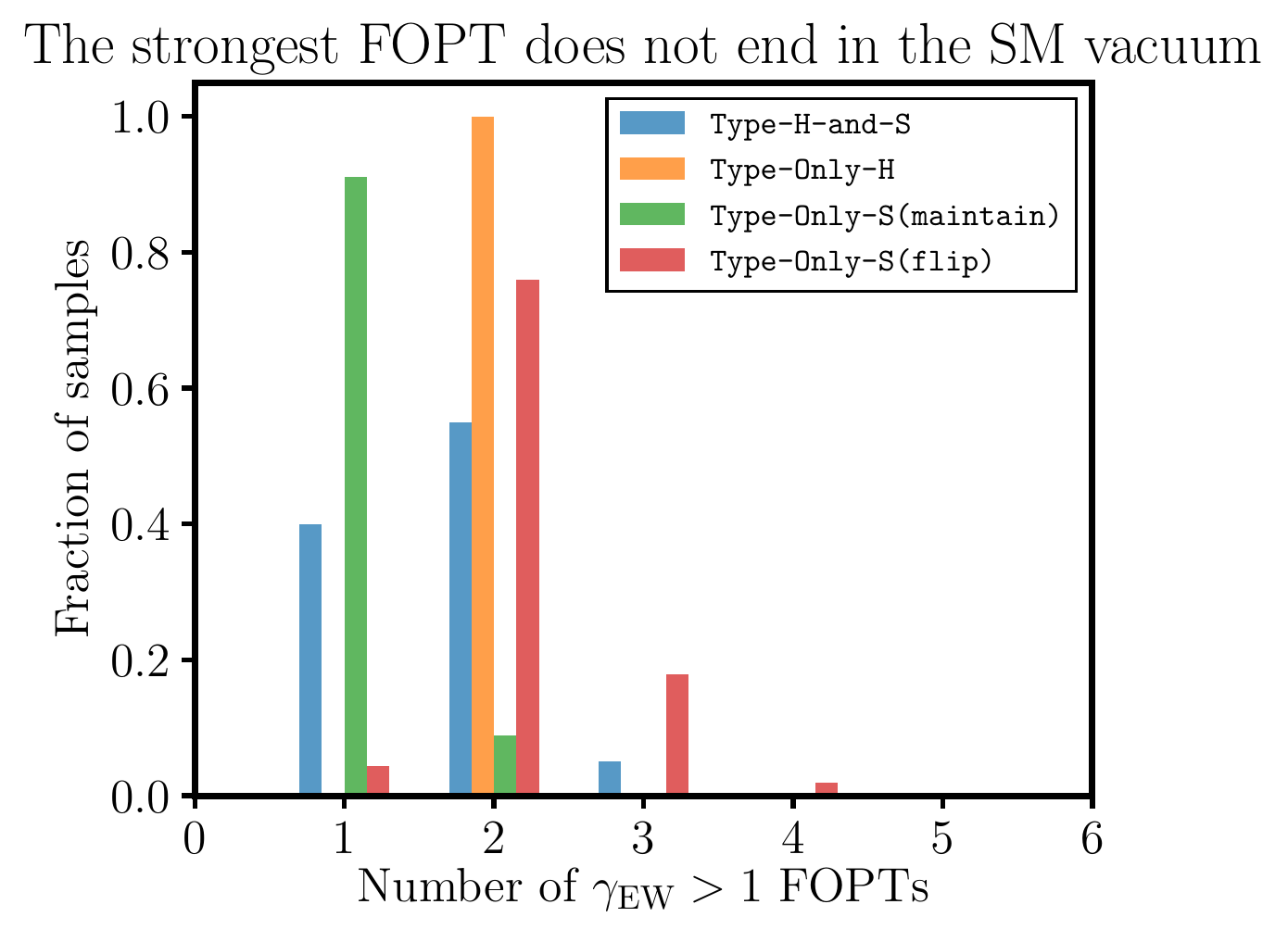}
\caption{Number of FOPTs with $\gammaEW \gtrsim 1$ per point, for points for which the strongest FOPT ends in the \globalmin (left panel) and does not end in the \globalmin (right panel).}
\label{fig:hist_nt}
\end{figure}

At first glance, these points might seem uninteresting, as they do not end in the correct zero-temperature vacuum. They may be especially interesting, however, as this means that in order for such samples to achieve the correct zero-temperature vacuum, there must be another EW FOPT transition or sequence of transitions that complete and end in the correct vacuum. Thus in \reffig{fig:hist_nt} we histogram the number of possible FOPTs with $\gammaEW \gtrsim 1$ for each sample.  Let us stress that strictly speaking, we count the number of temperatures at which two vacua are degenerate. This differs from the number of FOPTs that can take place in one cosmological history, since only particular routes through the phases are possible. For example the upper right panel of \reffig{fig:benchmarks_1} exhibits one or two FOPTs from phase 1 to phase 3, but we would count this though as three. Furthermore FOPTs may also occur between phases that were never degenerate, but such possibilities are not included in our count.  

For the samples that end in the \globalmin (left panel), there is usually a single FOPT with $\gammaEW>1$, except in the \TypeEW scenario, in which there are often two FOPTs with $\gammaEW>1$. For the samples that do not end in the \globalmin (right panel), almost all of \TypeEW and \TypeSs samples and about half the  \TypeEWS samples have more than one EW SFOPTs. We also checked that for most of them the second strongest FOPT {\it does} end in the \globalmin.

\begin{figure}[t]
\centering
{\includegraphics[width=.9\textwidth]{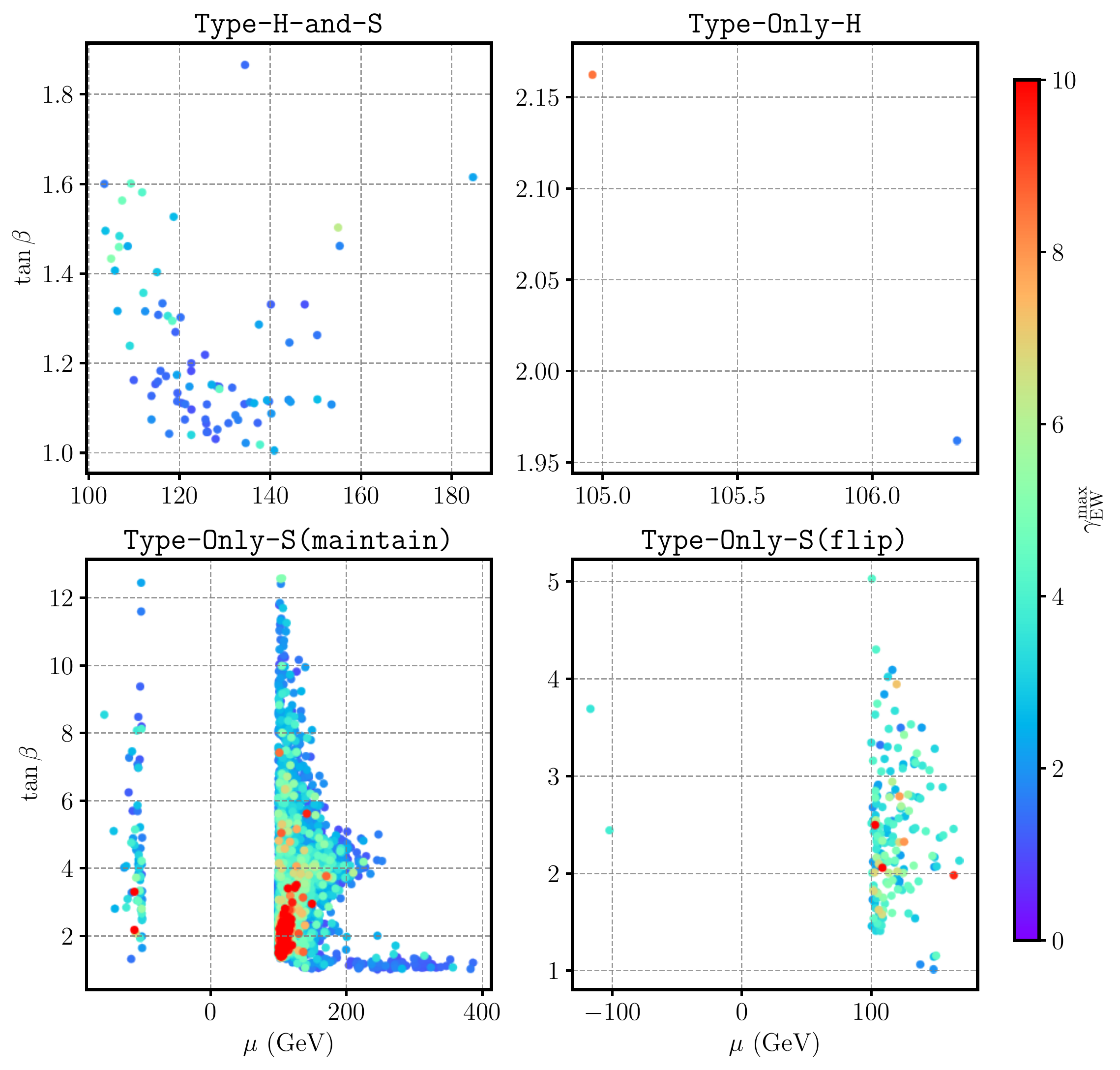}}
\caption{ $(\mueff,\tan\beta)$ for samples for which the strongest FOPT does not end in the \globalmin. The points are colored by the $\gammaEW$ of the strongest FOPT.}
\label{fig:mu_tanb:notDZTP}
\end{figure}

\begin{figure}[t]
\centering
{\includegraphics[width=.9\textwidth]{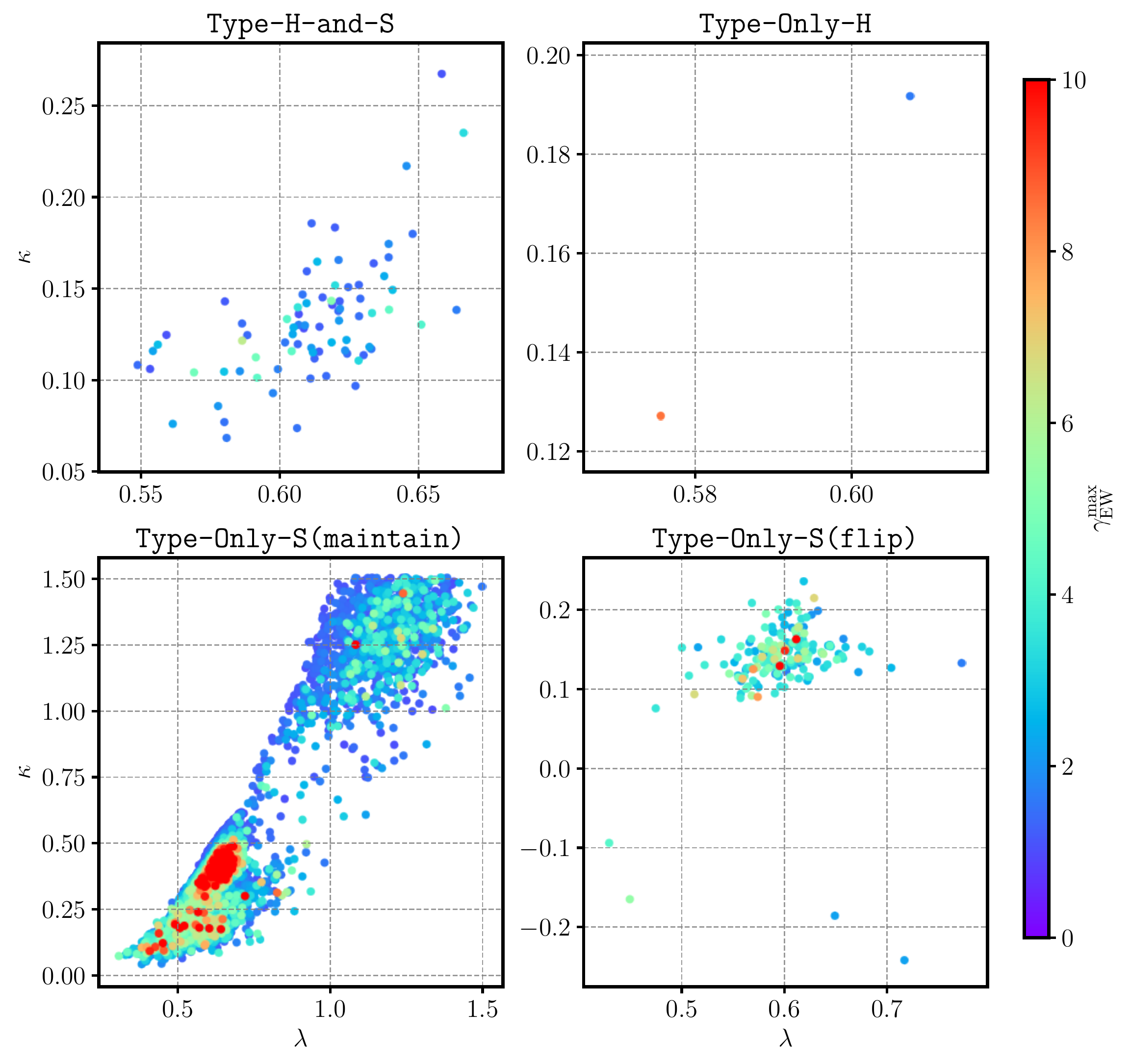}}
\caption{ $(\lambda,\kappa)$ for samples for which the strongest FOPT does not end in the \globalmin. The points are colored by the $\gammaEW$ of the strongest FOPT.}
\label{fig:lambda_kappa:notDZTP}
\end{figure}

\begin{figure}[t]
\centering
{\includegraphics[width=.9\textwidth]{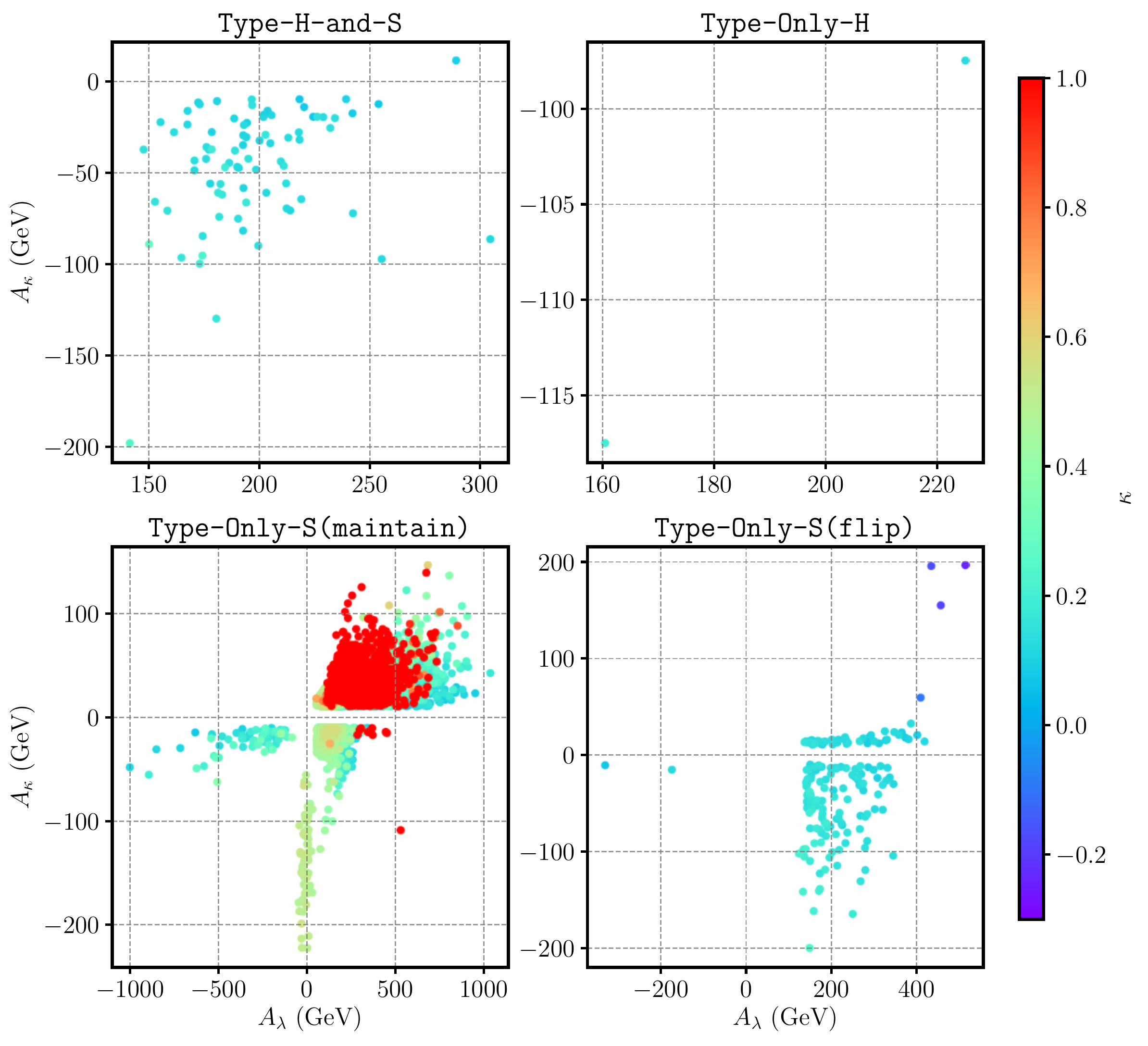}}
\caption{ $(A_\lambda,A_\kappa)$ for samples for which the strongest FOPT does not end in the \globalmin. The points are colored by the $\kappa$-parameter.}
\label{fig:Al_Ak:notDZTP}
\end{figure}

Thus, without further calculations, the samples for which the strongest FOPT does not end in the \globalmin could still potentially explain the observed baryon asymmetry. We display the parameter spaces in \reffig{fig:mu_tanb:notDZTP}, \reffig{fig:lambda_kappa:notDZTP} and \reffig{fig:Al_Ak:notDZTP}. Compared to the scenario in which the strongest FOPT ends in the \globalmin, the parameter spaces of \TypeEWS  and \TypeEW are roughly unchanged, while \TypeSs and \TypeSf exchange parameter spaces with each other. This is because here the \TypeSs (\TypeSf) requires a minimum on the singlet axis with $\minima{s}<0$ ($\minima{s}>0$), opposite to the \TypeSs (\TypeSf) scenarios in which the strongest FOPT ends in the \globalmin.

From \reffig{fig:mu_tanb:notDZTP} we see that the constraints on the effective $\mu$-parameter are stricter than they are in the scenario in which the strongest FOPT ends in the \globalmin, especially for small $\tan\beta$. The \TypeEWS, \TypeEW and \TypeSf scenarios require an effective $\mu$-parameter smaller than about 200\gev, whereas the \TypeSs permits $\mueff\lesssim 400\gev$. The slender bar in the \TypeSs scenario at $\tan\beta\simeq 1$ and $\mueff\in[200,400]\gev$ corresponds to samples with $T_C \gtrsim 200\gev$ for the strongest FOPT, displayed in the lower left panel of \reffig{fig:Tc_gamma:notDZTP}.

In \reffig{fig:lambda_kappa:notDZTP}, a visible difference appears in \TypeSs compared to \reffig{fig:lambda_kappa:DZTP}. The parameter space of $\lambda$ and $\kappa$ splits into two separate regions, and relatively large $\lambda\gtrsim 0.5$ is favored. For instance, when $\kappa \simeq 1.4$, here $\lambda$ is always larger than $1$, whereas in lower left panel of \reffig{fig:lambda_kappa:DZTP} $\lambda$ can be as low as $0.5$.

On the trilinear couplings $(A_\lambda,A_\kappa)$ plane, there are two additional regions in the \TypeSs scenario (lower left, \reffig{fig:Al_Ak:notDZTP}) compared to the \TypeSf samples for which the strongest FOPT end in the \globalmin (lower right, \reffig{fig:Al_Ak:DZTP}). First, there is an additional region at $A_\lambda \simeq 0\gev$. This region corresponds to the previously mentioned region at $\tan\beta\simeq 1$ and $\mueff\in[200,400]\gev$, with $T_C \gtrsim 200\gev$ for the strongest FOPT. Second, there is an additional region at $A_\kappa\simeq -50\gev$ and $A_\lambda > 0$. This region is similar to one in the \TypeSf scenario (lower right, \reffig{fig:Al_Ak:notDZTP}). Indeed, for this region, as well as the strongest FOPT that maintains the sign of singlet, there is another weaker FOPT that flips the sign of singlet.

In summary, the scenario in which the strongest FOPT does not end in
the \globalmin introduces new interesting regions of parameter space
that were not covered by the scenarios in which the strongest FOPT
ends in the \globalmin. These scenarios may be especially interesting
because they could be followed by additional FOPTs.  However, at the
same time there is an additional requirements to ensure that the
subsequent transitions actually lead to the EW breaking phase we
observe today, which we have not checked.

\subsection{Properties of the Higgs bosons}\label{sec:higgs_bosons}

As shown by our benchmark points, although our points pass experimental constraints from LEP and the LHC, our scenarios are not in a decoupling regime in which Higgses other than the $125\gev$ one are heavy.  This is not surprising since it is well known that in the NMSSM a light singlet Higgs state plays an important role in generating a FOPT that breaks EW symmetry~\cite{Profumo:2014opa, Kotwal:2016tex, Li:2019tfd}, without the need for light stops which are heavily constrained by LHC searches~\cite{Aaboud:2017ayj,CMS:2019qkm}. In fact, in all our benchmarks, all Higgs bosons are lighter than about $400\gev$, while there are always at least two CP even Higgs states with masses below $600\gev$ in the samples from our scan, with the SM-like Higgs being either $h_1$ or $h_2$.

\begin{figure}[t]
\centering
\includegraphics[width=.495\textwidth]{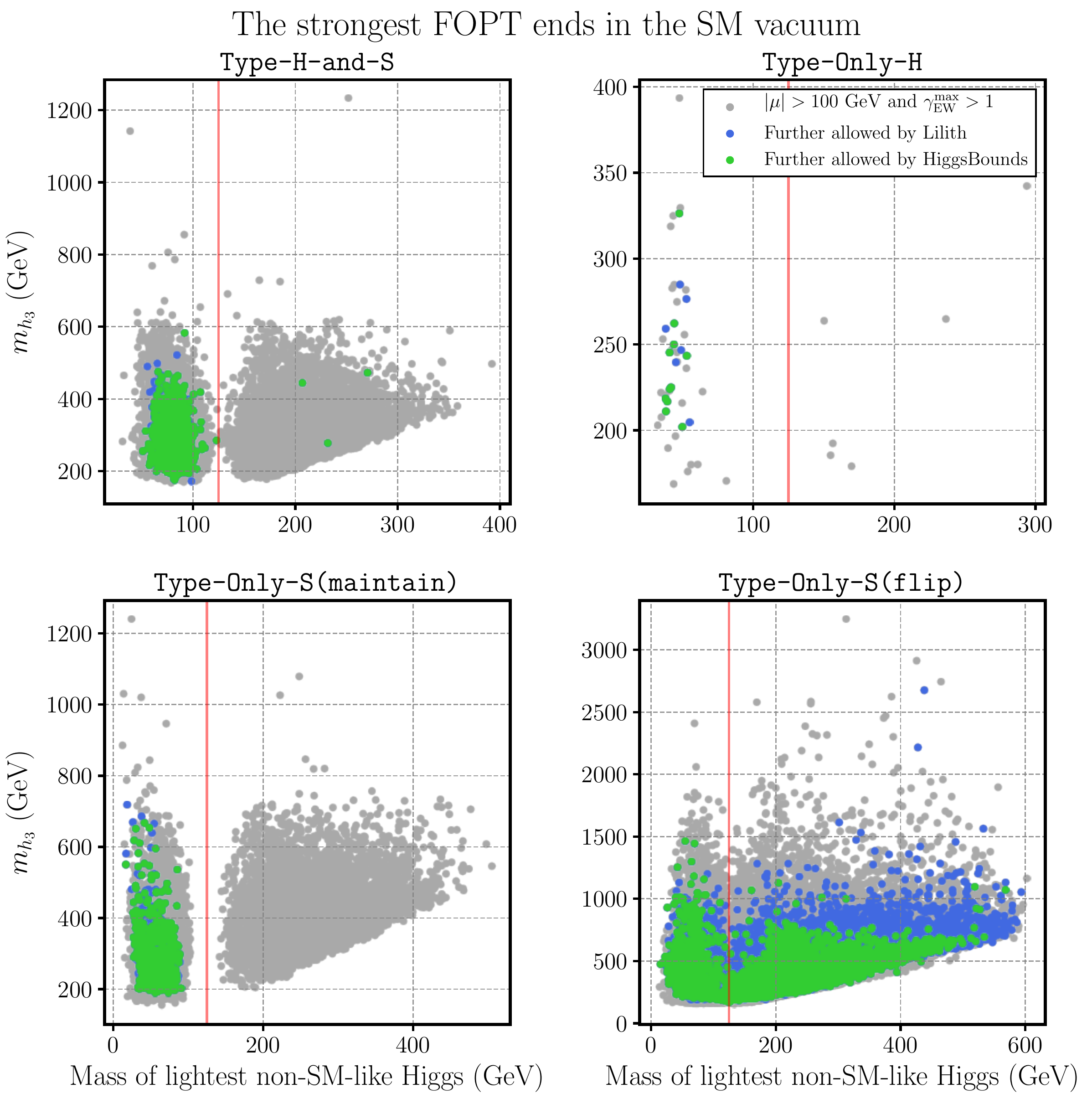}
\includegraphics[width=.495\textwidth]{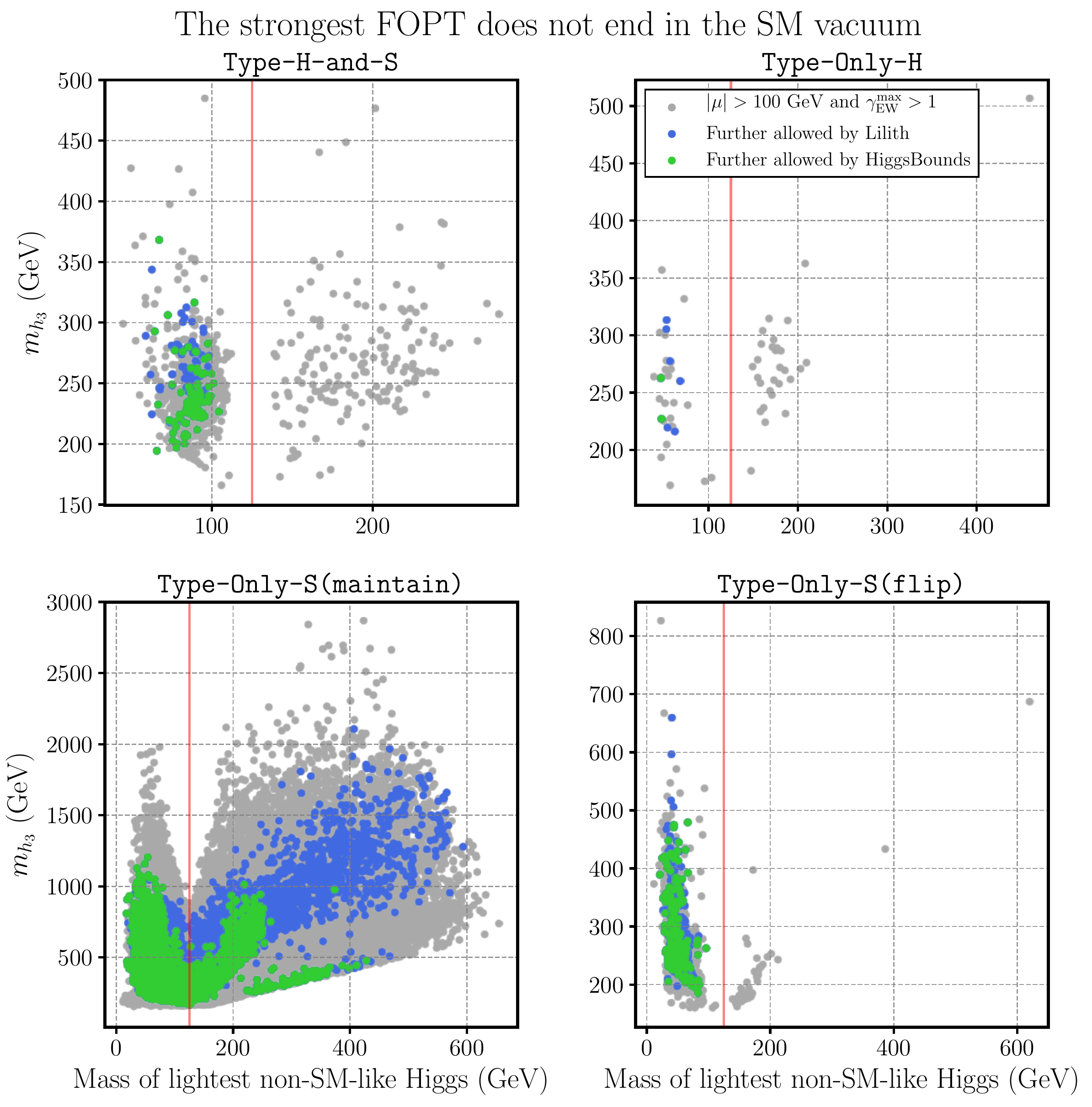}
\caption{Masses of the non-SM-like Higgs bosons in our four scenarios, for points for which the strongest FOPT ends in the \globalmin (left block of four plots) and does not end in the \globalmin (right block of four plots). We show points satisfying $\mu>100\gev$ and $\gammaEW>1$ (gray), further allowed by \texttt{Lilith-1.1.4\_DB-17.05}~\cite{Bernon:2015hsa} constraints on the SM-like  Higgs (blue), and by \texttt{HiggsBounds-5.3.2beta}~\cite{arXiv:0811.4169,arXiv:1102.1898,arXiv:1301.2345,arXiv:1311.0055,arXiv:1507.06706} constraints on non-SM-like Higgses (green). The vertical red line indicates $m_{h}=125\gev$ in each panel.}
\label{fig:higgs_masses}
\end{figure}

In \reffig{fig:higgs_masses} we show the masses of the non-SM-like CP even neutral Higgs bosons in our four scenarios by plotting the mass of $h_3$, which is never SM-like, against the mass of the Higgs (either $h_1$ or $h_2$) that did not play the role of the SM-like Higgs. Samples that are allowed by experimental constraints are shown by green points. We also show excluded samples to aid explanations (gray and blue points).

For the samples where the strongest FOPT ends in the \globalmin we see that the SM-like Higgs is actually the next to lightest CP even Higgs for almost all allowed samples (green points) in \TypeEWS, \TypeEW and \TypeSs, with just three exceptions that all appear in the \TypeEWS samples. In contrast, in the \TypeSf scenarios, the SM-like Higgs can be either the lightest Higgs or the next to lightest Higgs. The samples where the strongest FOPT does not end in the \globalmin show very similar results, but as usual the patterns of the \TypeSs and \TypeSf scenarios are exchanged. 

The reason we see so few samples where the SM-like Higgs is the lightest state for the categories mentioned above seems to be the constraints on the observed Higgs. We note that, although it is not clear in the plot, for these types of scenarios there are already a significantly larger number of gray points where the SM-like Higgs is the second lightest CP even Higgs boson than there are for the case where it is the lightest.  Applying the constraints on the SM-like Higgs from \texttt{Lilith-1.1.4\_DB-17.05} then reduces the number of samples where it is the lightest to almost zero.   The samples excluded by \texttt{HiggsBounds-5.3.2beta} in \TypeSf scenario for which the strongest FOPT ends in the \globalmin and \TypeSs scenario for which the strongest FOPT does not end in the \globalmin (shown by blue points) are mainly excluded by an LHC search for a scalar resonance decaying to a pair of $Z$ bosons~\cite{Sirunyan:2018qlb}. It is also worth noting that even without the requirement of  an SFOPT, similar observations have been made in the NMSSM previously. A  preference for the SM-like Higgs being the next to lightest one was also found in a global analysis of the NMSSM~\cite{AbdusSalam:2017uzr} that did not consider PTs.

Lastly, we note that many of the panels in \reffig{fig:higgs_masses} appear to indicate an upper bound on the mass of the heaviest Higgs, $m_{h_3}$, in each scenario. For example, for the \TypeEWS scenario in which the strongest FOPT ends in the \globalmin, the samples allowed by collider constraints on Higgs bosons (green points) stop at about $m_{h_3} \lesssim 500\gev$.
However, despite collecting more than three million samples,
we judged our coverage at the largest Higgs masses to be inadequate to reliably address the question of whether upper bounds on the Higgs masses exist,
as large masses may just be rare with our sampling strategy. We checked,
however, that experimental constraints on the Higgs sector appear to
be (at most) weakly sensitive to $m_{h_3}$. We thus anticipate that
there is in fact no upper bound on the mass of the heaviest Higgs, as we suspect
that it can be arbitrarily heavy without impacting the phase structure or Higgs observables.

\section{Conclusions}\label{sec:conclusions}

Motivated by EW baryogenesis and gravitational wave experiments, in this article we investigated the properties of PTs in the NMSSM. We employed an effective field theory approach to calculate the finite temperature effective potential by matching the NMSSM to the THDMS. By tracing the change in the minima of the effective potential with temperature, we mapped out the phase structure and computed the strengths of any EWPTs, $\gammaEW$. By scanning the parameter space of the NMSSM, we obtained millions of samples that featured an SFOPT with $\gammaEW>1$ and satisfied the constraints from LHC Higgs measurements and LEP bounds on charginos.

We classified them into three categories, \TypeEWS, \TypeEW and \TypeS, based on the  nature of the first PT in their cosmological histories. The \TypeS samples were further divided into \TypeSs and \TypeSf according to whether the singlet VEV changed sign during the strongest EWPT. In the \TypeEWS samples, the first PT in the cosmological history breaks EW symmetry and gives the singlet a VEV at the same time. This transition is usually the strongest one.

The \TypeEW samples, on the other hand, go through a series of PTs that break, restore and break again EW symmetry. The first one is a crossover transition during which only the $h_d$ field obtains a non-vanishing VEV, and the last one is the strongest EW FOPT. This scenario was by far the rarest in our scan. For the \TypeSs samples, during the first transition EW symmetry remains unbroken, but the singlet field obtains a non-vanishing VEV. Then EW symmetry breaks through a subsequent FOPT. Both of the transitions can be SFOPTs, which could give interesting gravitational wave signatures~\cite{Vieu:2018zze} as well as triggering an EW baryogenesis mechanism. The first  PT of the \TypeSf samples is usually a FOPT with very small $\gamma$, and the following strongest FOPT flips the sign of the singlet VEV. We found, however, that the tunneling rates in \TypeEW and \TypeSf scenarios could be problematic. For our benchmarks, the SFOPT in the \TypeEW scenario did not complete, and in the \TypeSf scenario, a preceding PT required to reach the SFOPT did not complete. Thus, unfortunately, these scenarios might not help EW baryogenesis.

The regions of NMSSM parameter space in which the four scenarios occur show different features. In the samples for which the strongest FOPT ends in the observed zero temperature phase:
\begin{itemize}
	\item The observed $125\gev$ Higgs is often the second lightest Higgs in the model, not the lightest one.
  \item All of the input parameters are severely constrained, except the SUSY scale $\msusy$ and stop trilinear $A_t$. 
  \item Quartic couplings of around $\lambda \simeq 0.6$ and $\kappa \simeq 0.2$ could result in an SFOPT in all our scenarios, though a broad range of couplings result in SFOPTs in the \TypeSf scenario, including couplings far away from limits on perturbativity.
  \item The scenarios predict different trilinear couplings, i.e., they are distinguishable on the $(A_\lambda,A_\kappa)$ plane. The $A_\lambda$ and $A_\kappa$ of the \TypeSf samples always have the same sign, while in the other scenarios the samples are concentrated in the quadrant of negative $A_\kappa$ and positive $A_\lambda$. Compared to the \TypeEWS scenario, the \TypeSs scenario favors smaller $|A_\kappa|$ and $A_\lambda$.
\end{itemize}

In addition we found substantial samples for which the strongest FOPT does not end in the \globalmin. The regions of parameter space are similar to the samples for which the strongest FOPT ends in the \globalmin, except that \TypeSs and \TypeSf exchange parameter spaces with each other. There are, furthermore, two additional regions that appear in the \TypeSs scenario, and one of them results in critical temperatures higher than 200\gev.

In summary, we mapped out and classified intricate patterns of symmetry breaking 
that are possible in the NMSSM, and checked which scenarios could in principle 
help provide a viable theory of EW baryogenesis or potentially lead to a gravitational wave signal. We found viable scenarios 
in which the Higgs fields and singlet or only singlet first acquired VEVs. 
We checked that the sequences of required PTs actually nucleated, contained a SFOPT, 
and that the model satisfied constraints from LEP and the LHC. The combination of 
constraints lead to the predictions that $\lambda \simeq 0.6$, $\kappa \simeq 0.2$ 
and that the observed $125\gev$ Higgs tends to be the second lightest Higgs in the model.

\section*{Acknowledgments}

We thank Sujeet Akula for early collaboration on the paper.
We are grateful for Margarete Mühlleitner and Jonathan Kozaczuk for responding to
queries regarding previous work and Junjie Cao for helpful remarks.
TRIUMF receives federal funding via a contribution agreement with the
National Research Council of Canada and the Natural Science and
Engineering Research Council of Canada. The work of P.A.\ is supported
by the Australian Research Council Future Fellowship grant
FT160100274. The work of C.B.\ was supported by the Australian
Research Council through the ARC Centre of Excellence for Particle
Physics at the Terascale (grant CE110001104). The work of P.A.\ and
C.B.\ are also supported with the Australian Research Council
Discovery Project grant DP180102209.
  
\appendix
\section{Field Dependent Masses}
\label{sec:Field_Dependent_Masses}
When exploring the potential away from minima we need to account for the Higgs field dependence of the $\overline{\text{MS}}$ mass eigenstates in~\eqref{eq:1LoopCorrection}. Therefore here we present the so-called field
dependent masses of the THDMS.

The field dependent masses of the gauge bosons and top, bottom and tau fermions are given by the simple tree-level expressions
\begin{equation}
\label{eq:GaugeBosonsTopBottomTauMasses}
\begin{gathered}
M_W^2 = \tfrac{1}{4} g^2 \left(\Huzr^2 + \Hdzr^2 \right), \quad
M_Z^2= \tfrac{1}{4}  \left( {g'}^2 + g^2 \right) \left(\Huzr^2 + \Hdzr^2 \right), \\
m_t = \tfrac{1}{\sqrt{2}} y_t \Huzr, \quad
m_b = \tfrac{1}{\sqrt{2}} y_b \Hdzr, \quad
m_{\tau} = \tfrac{1}{\sqrt{2}} y_{\tau} \Hdzr,
\end{gathered}
\end{equation}
where the gauge couplings are without GUT normalization,
and the $y_t$, $y_b$ and $y_\tau$ are the $(3,3)$ elements of the corresponding Yukawa matrices.

Since the Higgs states mix, the CP even, CP odd and charged
$\overline{\text{MS}}$ Higgs masses are the eigenvalues of the corresponding CP even, CP odd and charged mass
matrices. The mass matrix for the CP even neutral Higgs bosons, in the basis $\{H_d, H_u, S\}$, is
\begin{equation}
\begin{aligned}
 \left( M_{H^0}^2 \right)_{11} &=
\overline{m}_1^2
+ \tfrac32 \lambda_1 \Hdzr^2
+ \tfrac12 \lambda_5 \Sr^2
+ \tfrac12 (\lambda_3 + \lambda_4) \Huzr^2 ,
\\
 \left( M_{H^0}^2 \right)_{22} &=
\overline{m}_2^2
+ \tfrac32 \lambda_2 \Huzr^2
+ \tfrac12 \lambda_6 \Sr^2
+ \tfrac12 (\lambda_3 + \lambda_4) \Hdzr^2 ,
\\
\left( M_{H^0}^2 \right)_{33} &=
\overline{m}_{S}^2
- \sqrt{2} \re(m_5) {\Sr}
+ \tfrac12 \lambda_5 \Hdzr^2
+ \re(\lambda_7) \Huzr \Hdzr
+ \tfrac12 \lambda_6 \Huzr^2
+ 3 \lambda_8 \Sr^2 ,
\\
\left( M_{H^0}^2 \right)_{12} &= \left( M_{H^0}^2 \right)_{21} =
- \tfrac1{\sqrt{2}} \re(m_4) {\Sr}
+ \tfrac12 \re(\lambda_7) \Sr^2
+ (\lambda_3 + \lambda_4) \Huzr \Hdzr ,
\\
\left( M_{H^0}^2 \right)_{13} &= \left( M_{H^0}^2 \right)_{31} =
- \tfrac1{\sqrt{2}} \re(m_4) \Huzr
+ \lambda_5 \Hdzr {\Sr}
+ \re(\lambda_7) \Huzr {\Sr} ,
\\
\left( M_{H^0}^2 \right)_{23} &= \left( M_{H^0}^2 \right)_{32} =
- \tfrac1{\sqrt{2}} \re(m_4) \Hdzr
+ \re(\lambda_7) \Hdzr {\Sr}
+ \lambda_6 \Huzr {\Sr} .
\end{aligned}
\label{Eq:CpEvenHiggsMat}
\end{equation}
where we have written $\overline{m}_1^2$, $\overline{m}_2^2$ and $\overline{m}_{S}^2$ with a bar to denote the fact that in this context these are fixed to fulfill the following tree-level EW symmetry breaking (EWSB) conditions
\begin{equation}
\label{eq:TreeLevelSymmetryBreakingCondition}
\begin{aligned}
\overline{m}_1^2 &=
- \tfrac12 (\lambda_3 + \lambda_4) \vu^2
- \tfrac12 \lambda_1 \vd^2
- \tfrac12 \lambda_5 \vs^2
- \tfrac12 \re(\lambda_7) \frac{\vu \vs^2}{\vd}
+ \tfrac1{\sqrt{2}} \re(m_4) \frac{\vu \vs}{\vd},
\\
\overline{m}_2^2 &=
- \tfrac12 \lambda_2 \vu^2
- \tfrac12 (\lambda_3 +\lambda_4)\vd^2
- \tfrac12 \lambda_6 \vs^2
- \tfrac12 \re(\lambda_7) \frac{\vd \vs^2}{\vu}
+ \tfrac1{\sqrt{2}} \re(m_4) \frac{\vd \vs}{\vu},
\\
\overline{m}_{S}^2 &=
- \tfrac12 \lambda_6 \vu^2
- \tfrac12 \lambda_5 \vd^2
- \lambda_8 \vs^2
- \re(\lambda_7) \vd \vu
+ \tfrac1{\sqrt{2}} \re(m_4)\frac{\vu \vd}{\vs}
+ \tfrac1{\sqrt{2}} \re(m_5) \vs .
\end{aligned}
\end{equation}
Note that the VEVs appearing on the right hand side are the zero temperature VEVs, so $\overline{m}_1^2$, $\overline{m}_2^2$ and $\overline{m}_S^2$ do not vary with either temperature or with the fields. If we permit a complex phase in the THDMS parameters, there is
in fact an additional tadpole equation relating it to complex phases in the VEVs. As we assume real, CP conserving VEVs, however, this extra tadpole simply forces the complex phase in the THDMS parameters to vanish. The three CP even mass eigenstates, $h_1$, $h_2$ and $h_3$, are then found by diagonalizing $M^2_{H^0}$.

Similarly, the CP odd mass matrix is
\begin{equation}
\begin{aligned}
 \left( M_A^2 \right)_{11} &=
\overline{m}_1^2
+ \tfrac12 \lambda_1 \Hdzr^2
+ \tfrac12 \lambda_5 \Sr^2
+ \tfrac12 (\lambda_3 + \lambda_4) \Huzr^2 ,
\\
 \left( M_A^2 \right)_{22} &=
\overline{m}_2^2
+ \tfrac12 \lambda_2 \Huzr^2
+ \tfrac12 \lambda_6 \Sr^2
+ \tfrac12 (\lambda_3 + \lambda_4) \Hdzr^2 ,
\\
 \left( M_A^2 \right)_{33} &=
\overline{m}_S^2
+ \sqrt{2} \re(m_5) {\Sr}
+ \tfrac12 \lambda_5 \Hdzr^2
- \re(\lambda_7) \Huzr \Hdzr
+ \tfrac12 \lambda_6 \Huzr^2
+ \lambda_8 \Sr^2 ,
\\
\left( M_A^2 \right)_{12} &= \left( M_A^2 \right)_{21} =
\tfrac1{\sqrt{2}} \re(m_4) {\Sr}
- \tfrac12 \re(\lambda_7) \Sr^2 ,
\\
\left( M_A^2 \right)_{13}
&=
\left( M_A^2 \right)_{31}
=
\tfrac1{\sqrt{2}} \re(m_4) \Huzr
+ \re(\lambda_7) \Huzr {\Sr} ,
\\
\left( M_A^2 \right)_{23} &= \left( M_A^2 \right)_{32} =
\tfrac1{\sqrt{2}} \re(m_4) \Hdzr
+ \re(\lambda_7) \Hdzr {\Sr} .
\end{aligned}
\end{equation}
Diagonalizing it results in a neutral Goldstone boson $G^0$ and the two physical CP odd Higgs bosons, $A_1$ and $A_2$. The field dependent Goldstone masses are only zero at extrema of the tree-level potential.
Thus, away from extrema, we cannot easily distinguish Goldstone bosons from physical Higgs bosons. In the $\xi = 1$ gauge, however, they are treated on an equal footing and we do not need to identify Goldstones.

Finally, the charged Higgs mass matrix is
\begin{equation}
\begin{aligned}
 \left( M_{H^\pm}^2 \right)_{11} &=
\overline{m}_1^2
+ \tfrac12 \lambda_5 \Sr^2
+ \tfrac12 \lambda_1 \Hdzr^2
+ \tfrac12 \lambda_3 \Huzr^2 ,
\\
 \left( M_{H^\pm}^2 \right)_{22} &=
\overline{m}_2^2
+ \tfrac12 \lambda_6 \Sr^2
+ \tfrac12 \lambda_3 \Hdzr^2
+ \tfrac12 \lambda_2 \Huzr^2 ,
\\
\left( M_{H^\pm}^2 \right)_{21} &= \left( M_{H^\pm}^2 \right)_{12}^* =
\tfrac1{\sqrt{2}}  m_4 \Sr
- \tfrac12 \lambda_7 \Sr^2
- \tfrac12 \lambda_4 \Hdzr \Huzr .
\end{aligned}
\end{equation}
Diagonalizing it results in the charged Higgs boson, $H^\pm$ and the charged Goldstone boson $G^\pm$.

Gauge-fixing, however, alters the tree-level mass matrices, such that the field dependent scalar masses are gauge dependent. The CP even mass matrix receives no gauge-fixing contribution but the CP odd and charged mass matrices receive additional contributions in the $R_\xi$ gauge,
\begin{equation}\label{eq:GB_mass}
\begin{aligned}
\left( M_A^2 \right)_{11} &\to \left( M_A^2 \right)_{11}+ \frac{1}{4} \xi (g^2 + g'^2) \Hdzr^2,\\
\left( M_A^2 \right)_{12} &\to \left( M_A^2 \right)_{12}- \frac{1}{4} \xi (g^2 + g'^2) \Hdzr\Huzr,\\
\left( M_A^2 \right)_{22} &\to \left( M_A^2 \right)_{22}+ \frac{1}{4} \xi (g^2 + g'^2) \Huzr^2,\\
\left( M_{H^\pm}^2 \right)_{11} &\to \left( M_{H^\pm}^2 \right)_{11} + \frac{1}{4} \xi g^2 \Hdzr^2,\\
\left( M_{H^\pm}^2 \right)_{12} &\to \left( M_{H^\pm}^2 \right)_{12} - \frac{1}{4} \xi g^2 \Hdzr\Huzr,\\
\left( M_{H^\pm}^2 \right)_{22} &\to \left( M_{H^\pm}^2 \right)_{22} + \frac{1}{4} \xi g^2 \Huzr^2.
\end{aligned}
\end{equation}
The elements involving the singlet are unaffected. At the tree-level minimum, in which the Goldstone bosons are otherwise massless, the gauge-fixing contributions do not affect the masses of the physical Higgs bosons but result in Goldstone masses
\begin{equation}
\begin{aligned}
M_{G^0}^2 &= \xi M_Z^2,\\
M_{G^\pm}^2 &= \xi M_W^2,
\end{aligned}
\end{equation}
where $M_W$ and $M_Z$ are the masses of the $W$ and $Z$ bosons. 

\section{Numerical methods for FOPTs}\label{sec:pt_methods}

We first find all minima of the potential at $T=0$ and $T=1\tev$ to check that spontaneous symmetry breaking occurs\footnote{Our search for minima is restricted to field values within the range $-1.6\tev$ to $1.6\tev$.}, where in particular we reject points where the deepest $T=0$ minima is not the observed \globalmin. If it occurs, we trace the trajectory with temperature of every $T=0$ and $T=1\tev$ minima. We call the trajectory of a particular minima a phase (note though this definition cannot distinguish phases linked by second-order or crossover transitions). A phase ends at the temperature at which the minima disappears. If two phases coexist at the same temperature, there may exist a critical temperature at which they are degenerate.

We apply an algorithm developed in \cosmo~\cite{Wainwright:CosmoTransition} to trace phases in steps no greater than $\Delta T$:
\begin{enumerate}\setcounter{enumi}{-1}
    \item We select a minima $\minima{m} \equiv (\minima{h}_u, \minima{h}_d, \minima{s})$ at temperature $T$ to trace.

    \item\label{step:minimize} We use a local minimum finding algorithm, such as Nelder-Mead~\cite{Simplex}, to find the minimum
    $\minima{m}^\prime$ at $T^\prime = T  + \Delta T$.

    \item We check that the new minimum $\minima{m}^\prime$ lies close to that expected from a shift caused by thermal corrections.

    We calculate the difference
    \begin{equation}\label{eq:R}
    R = \max \left(
    \norm{\minima{m} + \left.\frac{\partial \minima{m}}{\partial T}\right|_{\minima{m}}\Delta T-\minima{m}^\prime },
    \norm{\minima{m}^\prime - \left.\frac{\partial \minima{m}}{\partial T}\right|_{\minima{m}^\prime}\Delta T -\minima{m}}
    \right).
    \end{equation}

    \item If $R \le \max R$, where $\max R$ governs the maximum acceptable changes in the field, we accept that the minima $\minima{m}^\prime$ at temperature $T^\prime$
     belongs to the same phase as the minima $\minima{m}$ at temperature $T$. We continue to trace the phase by returning to step~\ref{step:minimize} with $\minima{m} \to \minima{m}^\prime$ and $T \to T^\prime$, and we reset any changes to $\Delta T$.

     If $R > \max R$, we assume that the change in temperature dramatically changed the potential. We reduce the change in temperature by a factor of two, $\Delta T \to \Delta T/2$, and return to~\ref{step:minimize}.

     If, however, $R > \max R$ and $|\Delta T| < \min \Delta T$, where $\min \Delta T$ governs the smallest permissible step in temperature, we conclude that the phase must have ended, as the minima appears to change abruptly with a small change in temperature.
\end{enumerate}

We save the sequence of minima and temperature found through this process --- this is a phase. We find all the phases by tracing all $T=0$ minima up to at most $1\tev$ (the phase may end earlier) and all $T=1\tev$ minima down to $T=0$ (in which case $\Delta T < 0$). After removing degenerate phases, we denote the $i$-th phase by $\minima{m}_i(T)$.

If any two of the phases, e.g., the $i$-th and $j$-th phase, coexist between temperatures $T_1$ and $T_2$, and if
\begin{align}
V_\text{eff}(\minima{m}_i(T_1),T_1) &> V_\text{eff}(\minima{m}_j(T_1), T_1)\\
V_\text{eff}(\minima{m}_i(T_2),T_2) &< V_\text{eff}(\minima{m}_j(T_2), T_2)
\end{align}
there must exit a critical temperature, $T_C$, between temperatures $T_1$ and $T_2$ at which they are degenerate,
\begin{equation}
V_\text{eff}(\minima{m}_i(T_C),T_C) = V_\text{eff}(\minima{m}_j(T_C), T_C).
\end{equation}
We calculate the critical temperature using bisection, and find properties of the transition, e.g., the strength of transition from~\eqref{eq:gamma}.

\clearpage
\pagestyle{plain}

\bibliographystyle{JHEP}
\bibliography{bibliography}

\providecommand{\href}[2]{#2}\begingroup\raggedright\begin{thebibliography}{100}

\bibitem{Morrissey:2012db}
D.~E. Morrissey and M.~J. Ramsey-Musolf, \emph{{Electroweak baryogenesis}},
  \href{https://doi.org/10.1088/1367-2630/14/12/125003}{\emph{New J. Phys.}
  {\bfseries 14} (2012) 125003}
  [\href{https://arxiv.org/abs/1206.2942}{{\ttfamily 1206.2942}}].

\bibitem{Cline:2006ts}
J.~M. Cline, \emph{{Baryogenesis}},  in \emph{{Les Houches Summer School --
  Session 86: Particle Physics and Cosmology: The Fabric of Spacetime}}, 2006,
  \href{https://arxiv.org/abs/hep-ph/0609145}{{\ttfamily hep-ph/0609145}}.

\bibitem{White:2016bo}
G.~A. White, \emph{A Pedagogical Introduction to Electroweak Baryogenesis},
  2053-2571. Morgan \& Claypool Publishers, 2016,
  \href{https://doi.org/10.1088/978-1-6817-4457-5}{10.1088/978-1-6817-4457-5}.

\bibitem{Krnjaic:2016ycc}
G.~Krnjaic, \emph{{Can the Baryon Asymmetry Arise From Initial Conditions?}},
  \href{https://doi.org/10.1103/PhysRevD.96.035041}{\emph{Phys. Rev.}
  {\bfseries D96} (2017) 035041}
  [\href{https://arxiv.org/abs/1606.05344}{{\ttfamily 1606.05344}}].

\bibitem{Sakharov:1967dj}
A.~D. Sakharov, \emph{{Violation of $CP$ invariance, $C$ asymmetry, and baryon
  asymmetry of the universe}},
  \href{https://doi.org/10.1070/PU1991v034n05ABEH002497}{\emph{Pis'ma Zh. Eksp.
  Teor. Fiz.} {\bfseries 5} (1967) 32}.

\bibitem{DOnofrio:2015gop}
M.~D'Onofrio and K.~Rummukainen, \emph{{Standard model cross-over on the
  lattice}}, \href{https://doi.org/10.1103/PhysRevD.93.025003}{\emph{Phys.
  Rev.} {\bfseries D93} (2016) 025003}
  [\href{https://arxiv.org/abs/1508.07161}{{\ttfamily 1508.07161}}].

\bibitem{Patrignani:2016xqp}
{\scshape Particle Data Group} collaboration, \emph{{Review of Particle
  Physics}}, \href{https://doi.org/10.1088/1674-1137/40/10/100001}{\emph{Chin.
  Phys.} {\bfseries C40} (2016) 100001}.

\bibitem{Ade:2015xua}
{\scshape Planck} collaboration, \emph{{Planck 2015 results. XIII. Cosmological
  parameters}},
  \href{https://doi.org/10.1051/0004-6361/201525830}{\emph{Astron. Astrophys.}
  {\bfseries 594} (2016) A13}
  [\href{https://arxiv.org/abs/1502.01589}{{\ttfamily 1502.01589}}].

\bibitem{Trodden:1998ym}
M.~Trodden, \emph{{Electroweak baryogenesis}},
  \href{https://doi.org/10.1103/RevModPhys.71.1463}{\emph{Rev. Mod. Phys.}
  {\bfseries 71} (1999) 1463}
  [\href{https://arxiv.org/abs/hep-ph/9803479}{{\ttfamily hep-ph/9803479}}].

\bibitem{Cline:2008hr}
J.~M. Cline, M.~Jarvinen and F.~Sannino, \emph{{The Electroweak Phase
  Transition in Nearly Conformal Technicolor}},
  \href{https://doi.org/10.1103/PhysRevD.78.075027}{\emph{Phys. Rev.}
  {\bfseries D78} (2008) 075027}
  [\href{https://arxiv.org/abs/0808.1512}{{\ttfamily 0808.1512}}].

\bibitem{Cline:2009sn}
J.~M. Cline, G.~Laporte, H.~Yamashita and S.~Kraml, \emph{{Electroweak Phase
  Transition and LHC Signatures in the Singlet Majoron Model}},
  \href{https://doi.org/10.1088/1126-6708/2009/07/040}{\emph{JHEP} {\bfseries
  07} (2009) 040} [\href{https://arxiv.org/abs/0905.2559}{{\ttfamily
  0905.2559}}].

\bibitem{Borah:2012pu}
D.~Borah and J.~M. Cline, \emph{{Inert Doublet Dark Matter with Strong
  Electroweak Phase Transition}},
  \href{https://doi.org/10.1103/PhysRevD.86.055001}{\emph{Phys. Rev.}
  {\bfseries D86} (2012) 055001}
  [\href{https://arxiv.org/abs/1204.4722}{{\ttfamily 1204.4722}}].

\bibitem{Cline:2013bln}
J.~M. Cline and K.~Kainulainen, \emph{{Improved Electroweak Phase Transition
  with Subdominant Inert Doublet Dark Matter}},
  \href{https://doi.org/10.1103/PhysRevD.87.071701}{\emph{Phys. Rev.}
  {\bfseries D87} (2013) 071701}
  [\href{https://arxiv.org/abs/1302.2614}{{\ttfamily 1302.2614}}].

\bibitem{Konstandin:2013caa}
T.~Konstandin, \emph{{Quantum Transport and Electroweak Baryogenesis}},
  \href{https://doi.org/10.3367/UFNe.0183.201308a.0785}{\emph{Phys. Usp.}
  {\bfseries 56} (2013) 747} [\href{https://arxiv.org/abs/1302.6713}{{\ttfamily
  1302.6713}}].

\bibitem{Kozaczuk:2014kva}
J.~Kozaczuk, S.~Profumo, L.~S. Haskins and C.~L. Wainwright,
  \emph{{Cosmological phase transitions and their properties in the NMSSM}},
  \href{https://doi.org/10.1007/JHEP01(2015)144}{\emph{JHEP} {\bfseries 01}
  (2015) 144} [\href{https://arxiv.org/abs/1407.4134}{{\ttfamily 1407.4134}}].

\bibitem{Profumo:2014opa}
S.~Profumo, M.~J. Ramsey-Musolf, C.~L. Wainwright and P.~Winslow,
  \emph{{Singlet-catalyzed electroweak phase transitions and precision Higgs
  boson studies}},
  \href{https://doi.org/10.1103/PhysRevD.91.035018}{\emph{Phys. Rev.}
  {\bfseries D91} (2015) 035018}
  [\href{https://arxiv.org/abs/1407.5342}{{\ttfamily 1407.5342}}].

\bibitem{Curtin:2014jma}
D.~Curtin, P.~Meade and C.-T. Yu, \emph{{Testing Electroweak Baryogenesis with
  Future Colliders}},
  \href{https://doi.org/10.1007/JHEP11(2014)127}{\emph{JHEP} {\bfseries 11}
  (2014) 127} [\href{https://arxiv.org/abs/1409.0005}{{\ttfamily 1409.0005}}].

\bibitem{Huang:2015bta}
F.~P. Huang and C.~S. Li, \emph{{Electroweak baryogenesis in the framework of
  the effective field theory}},
  \href{https://doi.org/10.1103/PhysRevD.92.075014}{\emph{Phys. Rev.}
  {\bfseries D92} (2015) 075014}
  [\href{https://arxiv.org/abs/1507.08168}{{\ttfamily 1507.08168}}].

\bibitem{Inoue:2015pza}
S.~Inoue, G.~Ovanesyan and M.~J. Ramsey-Musolf, \emph{{Two-Step Electroweak
  Baryogenesis}}, \href{https://doi.org/10.1103/PhysRevD.93.015013}{\emph{Phys.
  Rev.} {\bfseries D93} (2016) 015013}
  [\href{https://arxiv.org/abs/1508.05404}{{\ttfamily 1508.05404}}].

\bibitem{Katz:2015uja}
A.~Katz, M.~Perelstein, M.~J. Ramsey-Musolf and P.~Winslow,
  \emph{{Stop-Catalyzed Baryogenesis Beyond the MSSM}},
  \href{https://doi.org/10.1103/PhysRevD.92.095019}{\emph{Phys. Rev.}
  {\bfseries D92} (2015) 095019}
  [\href{https://arxiv.org/abs/1509.02934}{{\ttfamily 1509.02934}}].

\bibitem{Fuyuto:2015ida}
K.~Fuyuto, J.~Hisano and E.~Senaha, \emph{{Toward verification of electroweak
  baryogenesis by electric dipole moments}},
  \href{https://doi.org/10.1016/j.physletb.2016.02.053}{\emph{Phys. Lett.}
  {\bfseries B755} (2016) 491}
  [\href{https://arxiv.org/abs/1510.04485}{{\ttfamily 1510.04485}}].

\bibitem{Huang:2015izx}
F.~P. Huang, P.-H. Gu, P.-F. Yin, Z.-H. Yu and X.~Zhang, \emph{{Testing the
  electroweak phase transition and electroweak baryogenesis at the LHC and a
  circular electron-positron collider}},
  \href{https://doi.org/10.1103/PhysRevD.93.103515}{\emph{Phys. Rev.}
  {\bfseries D93} (2016) 103515}
  [\href{https://arxiv.org/abs/1511.03969}{{\ttfamily 1511.03969}}].

\bibitem{Kobakhidze:2015xlz}
A.~Kobakhidze, L.~Wu and J.~Yue, \emph{{Electroweak Baryogenesis with Anomalous
  Higgs Couplings}}, \href{https://doi.org/10.1007/JHEP04(2016)011}{\emph{JHEP}
  {\bfseries 04} (2016) 011}
  [\href{https://arxiv.org/abs/1512.08922}{{\ttfamily 1512.08922}}].

\bibitem{Huang:2016odd}
F.~P. Huang, Y.~Wan, D.-G. Wang, Y.-F. Cai and X.~Zhang, \emph{{Hearing the
  echoes of electroweak baryogenesis with gravitational wave detectors}},
  \href{https://doi.org/10.1103/PhysRevD.94.041702}{\emph{Phys. Rev.}
  {\bfseries D94} (2016) 041702}
  [\href{https://arxiv.org/abs/1601.01640}{{\ttfamily 1601.01640}}].

\bibitem{Kotwal:2016tex}
A.~V. Kotwal, M.~J. Ramsey-Musolf, J.~M. No and P.~Winslow,
  \emph{{Singlet-catalyzed electroweak phase transitions in the 100 TeV
  frontier}}, \href{https://doi.org/10.1103/PhysRevD.94.035022}{\emph{Phys.
  Rev.} {\bfseries D94} (2016) 035022}
  [\href{https://arxiv.org/abs/1605.06123}{{\ttfamily 1605.06123}}].

\bibitem{Vaskonen:2016yiu}
V.~Vaskonen, \emph{{Electroweak baryogenesis and gravitational waves from a
  real scalar singlet}},
  \href{https://doi.org/10.1103/PhysRevD.95.123515}{\emph{Phys. Rev.}
  {\bfseries D95} (2017) 123515}
  [\href{https://arxiv.org/abs/1611.02073}{{\ttfamily 1611.02073}}].

\bibitem{Balazs:2016yvi}
C.~Balazs, G.~White and J.~Yue, \emph{{Effective field theory, electric dipole
  moments and electroweak baryogenesis}},
  \href{https://doi.org/10.1007/JHEP03(2017)030}{\emph{JHEP} {\bfseries 03}
  (2017) 030} [\href{https://arxiv.org/abs/1612.01270}{{\ttfamily
  1612.01270}}].

\bibitem{Beniwal:2017eik}
A.~Beniwal, M.~Lewicki, J.~D. Wells, M.~White and A.~G. Williams,
  \emph{{Gravitational wave, collider and dark matter signals from a scalar
  singlet electroweak baryogenesis}},
  \href{https://doi.org/10.1007/JHEP08(2017)108}{\emph{JHEP} {\bfseries 08}
  (2017) 108} [\href{https://arxiv.org/abs/1702.06124}{{\ttfamily
  1702.06124}}].

\bibitem{Kurup:2017dzf}
G.~Kurup and M.~Perelstein, \emph{{Dynamics of Electroweak Phase Transition In
  Singlet-Scalar Extension of the Standard Model}},
  \href{https://doi.org/10.1103/PhysRevD.96.015036}{\emph{Phys. Rev.}
  {\bfseries D96} (2017) 015036}
  [\href{https://arxiv.org/abs/1704.03381}{{\ttfamily 1704.03381}}].

\bibitem{Akula:2017yfr}
S.~Akula, C.~Balázs, L.~Dunn and G.~White, \emph{{Electroweak baryogenesis in
  the $ {\mathbb{Z}}_3 $ -invariant NMSSM}},
  \href{https://doi.org/10.1007/JHEP11(2017)051}{\emph{JHEP} {\bfseries 11}
  (2017) 051} [\href{https://arxiv.org/abs/1706.09898}{{\ttfamily
  1706.09898}}].

\bibitem{Chiang:2017nmu}
C.-W. Chiang, M.~J. Ramsey-Musolf and E.~Senaha, \emph{{Standard Model with a
  Complex Scalar Singlet: Cosmological Implications and Theoretical
  Considerations}},
  \href{https://doi.org/10.1103/PhysRevD.97.015005}{\emph{Phys. Rev.}
  {\bfseries D97} (2018) 015005}
  [\href{https://arxiv.org/abs/1707.09960}{{\ttfamily 1707.09960}}].

\bibitem{Cao:2017oez}
Q.-H. Cao, F.~P. Huang, K.-P. Xie and X.~Zhang, \emph{{Testing the electroweak
  phase transition in scalar extension models at lepton colliders}},
  \href{https://doi.org/10.1088/1674-1137/42/2/023103}{\emph{Chin. Phys.}
  {\bfseries C42} (2018) 023103}
  [\href{https://arxiv.org/abs/1708.04737}{{\ttfamily 1708.04737}}].

\bibitem{Ramsey-Musolf:2017tgh}
M.~J. Ramsey-Musolf, P.~Winslow and G.~White, \emph{{Color Breaking
  Baryogenesis}}, \href{https://doi.org/10.1103/PhysRevD.97.123509}{\emph{Phys.
  Rev.} {\bfseries D97} (2018) 123509}
  [\href{https://arxiv.org/abs/1708.07511}{{\ttfamily 1708.07511}}].

\bibitem{Huang:2017kzu}
F.~P. Huang and C.~S. Li, \emph{{Probing the baryogenesis and dark matter
  relaxed in phase transition by gravitational waves and colliders}},
  \href{https://doi.org/10.1103/PhysRevD.96.095028}{\emph{Phys. Rev.}
  {\bfseries D96} (2017) 095028}
  [\href{https://arxiv.org/abs/1709.09691}{{\ttfamily 1709.09691}}].

\bibitem{deVries:2017ncy}
J.~de~Vries, M.~Postma, J.~van~de Vis and G.~White, \emph{{Electroweak
  Baryogenesis and the Standard Model Effective Field Theory}},
  \href{https://doi.org/10.1007/JHEP01(2018)089}{\emph{JHEP} {\bfseries 01}
  (2018) 089} [\href{https://arxiv.org/abs/1710.04061}{{\ttfamily
  1710.04061}}].

\bibitem{Niemi:2018asa}
L.~Niemi, H.~H. Patel, M.~J. Ramsey-Musolf, T.~V.~I. Tenkanen and D.~J. Weir,
  \emph{{Electroweak phase transition in the $\Sigma$SM - I: Dimensional
  reduction}},  \href{https://arxiv.org/abs/1802.10500}{{\ttfamily
  1802.10500}}.

\bibitem{Modak:2018csw}
T.~Modak and E.~Senaha, \emph{{Electroweak baryogenesis via bottom transport}},
  \href{https://doi.org/10.1103/PhysRevD.99.115022}{\emph{Phys. Rev.}
  {\bfseries D99} (2019) 115022}
  [\href{https://arxiv.org/abs/1811.08088}{{\ttfamily 1811.08088}}].

\bibitem{Carena:2018cjh}
M.~Carena, M.~Quirós and Y.~Zhang, \emph{{Electroweak Baryogenesis from
  Dark-Sector CP Violation}},
  \href{https://doi.org/10.1103/PhysRevLett.122.201802}{\emph{Phys. Rev. Lett.}
  {\bfseries 122} (2019) 201802}
  [\href{https://arxiv.org/abs/1811.09719}{{\ttfamily 1811.09719}}].

\bibitem{Chala:2018opy}
M.~Chala, M.~Ramos and M.~Spannowsky, \emph{{Gravitational wave and collider
  probes of a triplet Higgs sector with a low cutoff}},
  \href{https://doi.org/10.1140/epjc/s10052-019-6655-1}{\emph{Eur. Phys. J.}
  {\bfseries C79} (2019) 156}
  [\href{https://arxiv.org/abs/1812.01901}{{\ttfamily 1812.01901}}].

\bibitem{Zhou:2018zli}
R.~Zhou, W.~Cheng, X.~Deng, L.~Bian and Y.~Wu, \emph{{Electroweak phase
  transition and Higgs phenomenology in the Georgi-Machacek model}},
  \href{https://doi.org/10.1007/JHEP01(2019)216}{\emph{JHEP} {\bfseries 01}
  (2019) 216} [\href{https://arxiv.org/abs/1812.06217}{{\ttfamily
  1812.06217}}].

\bibitem{Alves:2018jsw}
A.~Alves, T.~Ghosh, H.-K. Guo, K.~Sinha and D.~Vagie, \emph{{Collider and
  Gravitational Wave Complementarity in Exploring the Singlet Extension of the
  Standard Model}}, \href{https://doi.org/10.1007/JHEP04(2019)052}{\emph{JHEP}
  {\bfseries 04} (2019) 052}
  [\href{https://arxiv.org/abs/1812.09333}{{\ttfamily 1812.09333}}].

\bibitem{YaserAyazi:2019caf}
S.~Yaser~Ayazi and A.~Mohamadnejad, \emph{{Conformal vector dark matter and
  strongly first-order electroweak phase transition}},
  \href{https://doi.org/10.1007/JHEP03(2019)181}{\emph{JHEP} {\bfseries 03}
  (2019) 181} [\href{https://arxiv.org/abs/1901.04168}{{\ttfamily
  1901.04168}}].

\bibitem{Mohamadnejad:2019vzg}
A.~Mohamadnejad, \emph{{Gravitational waves from scale-invariant vector dark
  matter model: Probing below the neutrino-floor}},
  \href{https://arxiv.org/abs/1907.08899}{{\ttfamily 1907.08899}}.

\bibitem{Witten:1984rs}
E.~Witten, \emph{{Cosmic Separation of Phases}},
  \href{https://doi.org/10.1103/PhysRevD.30.272}{\emph{Phys. Rev.} {\bfseries
  D30} (1984) 272}.

\bibitem{1986MNRAS.218..629H}
C.~J. {Hogan}, \emph{{Gravitational radiation from cosmological phase
  transitions}}, \href{https://doi.org/10.1093/mnras/218.4.629}{\emph{Monthly
  Notices of the Royal Astronomical Society} {\bfseries 218} (1986) 629}.

\bibitem{Krauss:1991qu}
L.~M. Krauss, \emph{{Gravitational waves from global phase transitions}},
  \href{https://doi.org/10.1016/0370-2693(92)90425-4}{\emph{Phys. Lett.}
  {\bfseries B284} (1992) 229}.

\bibitem{Kosowsky:1991ua}
A.~Kosowsky, M.~S. Turner and R.~Watkins, \emph{{Gravitational radiation from
  colliding vacuum bubbles}},
  \href{https://doi.org/10.1103/PhysRevD.45.4514}{\emph{Phys. Rev.} {\bfseries
  D45} (1992) 4514}.

\bibitem{Kosowsky:1992rz}
A.~Kosowsky, M.~S. Turner and R.~Watkins, \emph{{Gravitational waves from first
  order cosmological phase transitions}},
  \href{https://doi.org/10.1103/PhysRevLett.69.2026}{\emph{Phys. Rev. Lett.}
  {\bfseries 69} (1992) 2026}.

\bibitem{Kamionkowski:1993fg}
M.~Kamionkowski, A.~Kosowsky and M.~S. Turner, \emph{{Gravitational radiation
  from first order phase transitions}},
  \href{https://doi.org/10.1103/PhysRevD.49.2837}{\emph{Phys. Rev.} {\bfseries
  D49} (1994) 2837} [\href{https://arxiv.org/abs/astro-ph/9310044}{{\ttfamily
  astro-ph/9310044}}].

\bibitem{Lee:2004we}
C.~Lee, V.~Cirigliano and M.~J. Ramsey-Musolf, \emph{{Resonant relaxation in
  electroweak baryogenesis}},
  \href{https://doi.org/10.1103/PhysRevD.71.075010}{\emph{Phys. Rev. D}
  {\bfseries 71} (2005) 075010}
  [\href{https://arxiv.org/abs/hep-ph/0412354}{{\ttfamily hep-ph/0412354}}].

\bibitem{Balazs:2004ae}
C.~Bal\'azs, M.~Carena, A.~Menon, D.~E. Morrissey and C.~E.~M. Wagner,
  \emph{{Supersymmetric origin of matter}},
  \href{https://doi.org/10.1103/PhysRevD.71.075002}{\emph{Phys. Rev. D}
  {\bfseries 71} (2005) 075002}
  [\href{https://arxiv.org/abs/hep-ph/0412264}{{\ttfamily hep-ph/0412264}}].

\bibitem{Liebler:2015ddv}
S.~Liebler, S.~Profumo and T.~Stefaniak, \emph{{Light Stop Mass Limits from
  Higgs Rate Measurements in the MSSM: Is MSSM Electroweak Baryogenesis Still
  Alive After All?}},
  \href{https://doi.org/10.1007/JHEP04(2016)143}{\emph{JHEP} {\bfseries 04}
  (2016) 143} [\href{https://arxiv.org/abs/1512.09172}{{\ttfamily
  1512.09172}}].

\bibitem{Maniatis:2009re}
M.~Maniatis, \emph{{The Next-to-Minimal Supersymmetric extension of the
  Standard Model reviewed}},
  \href{https://doi.org/10.1142/S0217751X10049827}{\emph{Int. J. Mod. Phys.}
  {\bfseries A25} (2010) 3505}
  [\href{https://arxiv.org/abs/0906.0777}{{\ttfamily 0906.0777}}].

\bibitem{Ellwanger:2009dp}
U.~Ellwanger, C.~Hugonie and A.~M. Teixeira, \emph{{The Next-to-Minimal
  Supersymmetric Standard Model}},
  \href{https://doi.org/10.1016/j.physrep.2010.07.001}{\emph{Phys. Rept.}
  {\bfseries 496} (2010) 1} [\href{https://arxiv.org/abs/0910.1785}{{\ttfamily
  0910.1785}}].

\bibitem{Li:2019tfd}
H.-L. Li, M.~Ramsey-Musolf and S.~Willocq, \emph{{Probing a Scalar
  Singlet-Catalyzed Electroweak Phase Transition with Resonant Di-Higgs
  Production in the $4b$ Channel}},
  \href{https://arxiv.org/abs/1906.05289}{{\ttfamily 1906.05289}}.

\bibitem{Bian:2017wfv}
L.~Bian, H.-K. Guo and J.~Shu, \emph{{Gravitational Waves, baryon asymmetry of
  the universe and electric dipole moment in the CP-violating NMSSM}},
  {\emph{Chin. Phys.} {\bfseries C42} (2018) 093106}
  [\href{https://arxiv.org/abs/1704.02488}{{\ttfamily 1704.02488}}].

\bibitem{Huber:2006wf}
S.~J. Huber, T.~Konstandin, T.~Prokopec and M.~G. Schmidt, \emph{{Electroweak
  Phase Transition and Baryogenesis in the nMSSM}},
  \href{https://doi.org/10.1016/j.nuclphysb.2006.09.003}{\emph{Nucl. Phys. B}
  {\bfseries 757} (2006) 172}
  [\href{https://arxiv.org/abs/hep-ph/0606298}{{\ttfamily hep-ph/0606298}}].

\bibitem{Balazs:2013cia}
{\relax Cs}.~Bal\'azs, A.~Mazumdar, E.~Pukartas and G.~White,
  \emph{{Baryogenesis, dark matter and inflation in the next-to-minimal
  supersymmetric standard model}},
  \href{https://doi.org/10.1007/JHEP01(2014)073}{\emph{JHEP} {\bfseries 01}
  (2014) 073} [\href{https://arxiv.org/abs/1309.5091}{{\ttfamily 1309.5091}}].

\bibitem{Cheung:2012pg}
K.~Cheung, T.-J. Hou, J.~S. Lee and E.~Senaha, \emph{{Singlino-driven
  Electroweak Baryogenesis in the Next-to-MSSM}},
  \href{https://doi.org/10.1016/j.physletb.2012.02.070}{\emph{Phys. Lett. B}
  {\bfseries 710} (2012) 188}
  [\href{https://arxiv.org/abs/1201.3781}{{\ttfamily 1201.3781}}].

\bibitem{Huang:2014ifa}
W.~Huang, Z.~Kang, J.~Shu, P.~Wu and J.~M. Yang, \emph{{New insights in the
  electroweak phase transition in the NMSSM}},
  \href{https://doi.org/10.1103/PhysRevD.91.025006}{\emph{Phys. Rev.}
  {\bfseries D91} (2015) 025006}
  [\href{https://arxiv.org/abs/1405.1152}{{\ttfamily 1405.1152}}].

\bibitem{Demidov:2016wcv}
S.~V. Demidov, D.~S. Gorbunov and D.~V. Kirpichnikov, \emph{{Split NMSSM with
  electroweak baryogenesis}}, \href{https://doi.org/10.1007/JHEP11(2016)148,
  10.1007/JHEP08(2017)080}{\emph{JHEP} {\bfseries 11} (2016) 148}
  [\href{https://arxiv.org/abs/1608.01985}{{\ttfamily 1608.01985}}].

\bibitem{Bi:2015qva}
X.-J. Bi, L.~Bian, W.~Huang, J.~Shu and P.-F. Yin, \emph{{Interpretation of the
  Galactic Center excess and electroweak phase transition in the NMSSM}},
  \href{https://doi.org/10.1103/PhysRevD.92.023507}{\emph{Phys. Rev.}
  {\bfseries D92} (2015) 023507}
  [\href{https://arxiv.org/abs/1503.03749}{{\ttfamily 1503.03749}}].

\bibitem{Carena:2011jy}
M.~Carena, N.~R. Shah and C.~E.~M. Wagner, \emph{{Light Dark Matter and the
  Electroweak Phase Transition in the NMSSM}},
  \href{https://doi.org/10.1103/PhysRevD.85.036003}{\emph{Phys. Rev.}
  {\bfseries D85} (2012) 036003}
  [\href{https://arxiv.org/abs/1110.4378}{{\ttfamily 1110.4378}}].

\bibitem{Menon:2004wv}
A.~Menon, D.~E. Morrissey and C.~E.~M. Wagner, \emph{{Electroweak baryogenesis
  and dark matter in the nMSSM}},
  \href{https://doi.org/10.1103/PhysRevD.70.035005}{\emph{Phys. Rev.}
  {\bfseries D70} (2004) 035005}
  [\href{https://arxiv.org/abs/hep-ph/0404184}{{\ttfamily hep-ph/0404184}}].

\bibitem{Bell:2019mbn}
N.~F. Bell, M.~J. Dolan, L.~S. Friedrich, M.~J. Ramsey-Musolf and R.~R. Volkas,
  \emph{{Electroweak Baryogenesis with Vector-like Leptons and Scalar
  Singlets}},  \href{https://arxiv.org/abs/1903.11255}{{\ttfamily 1903.11255}}.

\bibitem{deVries:2018tgs}
J.~De~Vries, M.~Postma and J.~van~de Vis, \emph{{The role of leptons in
  electroweak baryogenesis}},
  \href{https://doi.org/10.1007/JHEP04(2019)024}{\emph{JHEP} {\bfseries 04}
  (2019) 024} [\href{https://arxiv.org/abs/1811.11104}{{\ttfamily
  1811.11104}}].

\bibitem{Chen:2017com}
C.-Y. Chen, H.-L. Li and M.~Ramsey-Musolf, \emph{{CP-Violation in the Two Higgs
  Doublet Model: from the LHC to EDMs}},
  \href{https://doi.org/10.1103/PhysRevD.97.015020}{\emph{Phys. Rev.}
  {\bfseries D97} (2018) 015020}
  [\href{https://arxiv.org/abs/1708.00435}{{\ttfamily 1708.00435}}].

\bibitem{Guo:2016ixx}
H.-K. Guo, Y.-Y. Li, T.~Liu, M.~Ramsey-Musolf and J.~Shu,
  \emph{{Lepton-Flavored Electroweak Baryogenesis}},
  \href{https://doi.org/10.1103/PhysRevD.96.115034}{\emph{Phys. Rev.}
  {\bfseries D96} (2017) 115034}
  [\href{https://arxiv.org/abs/1609.09849}{{\ttfamily 1609.09849}}].

\bibitem{Chao:2015uoa}
W.~Chao and M.~J. Ramsey-Musolf, \emph{{Catalysis of Electroweak Baryogenesis
  via Fermionic Higgs Portal Dark Matter}},
  \href{https://arxiv.org/abs/1503.00028}{{\ttfamily 1503.00028}}.

\bibitem{Cline:2017qpe}
J.~M. Cline, K.~Kainulainen and D.~Tucker-Smith, \emph{{Electroweak
  baryogenesis from a dark sector}},
  \href{https://doi.org/10.1103/PhysRevD.95.115006}{\emph{Phys. Rev.}
  {\bfseries D95} (2017) 115006}
  [\href{https://arxiv.org/abs/1702.08909}{{\ttfamily 1702.08909}}].

\bibitem{Cline:2012hg}
J.~M. Cline and K.~Kainulainen, \emph{{Electroweak baryogenesis and dark matter
  from a singlet Higgs}},
  \href{https://doi.org/10.1088/1475-7516/2013/01/012}{\emph{JCAP} {\bfseries
  1301} (2013) 012} [\href{https://arxiv.org/abs/1210.4196}{{\ttfamily
  1210.4196}}].

\bibitem{Cline:2011mm}
J.~M. Cline, K.~Kainulainen and M.~Trott, \emph{{Electroweak Baryogenesis in
  Two Higgs Doublet Models and B meson anomalies}},
  \href{https://doi.org/10.1007/JHEP11(2011)089}{\emph{JHEP} {\bfseries 11}
  (2011) 089} [\href{https://arxiv.org/abs/1107.3559}{{\ttfamily 1107.3559}}].

\bibitem{Carena:2018vpt}
M.~Carena, Z.~Liu and M.~Riembau, \emph{{Probing the electroweak phase
  transition via enhanced di-Higgs boson production}},
  \href{https://doi.org/10.1103/PhysRevD.97.095032}{\emph{Phys. Rev.}
  {\bfseries D97} (2018) 095032}
  [\href{https://arxiv.org/abs/1801.00794}{{\ttfamily 1801.00794}}].

\bibitem{Cline:1997vk}
J.~M. Cline, M.~Joyce and K.~Kainulainen, \emph{{Supersymmetric electroweak
  baryogenesis in the WKB approximation}},
  \href{https://doi.org/10.1016/S0370-2693(97)01361-0}{\emph{Phys. Lett. B}
  {\bfseries 417} (1998) 79}
  [\href{https://arxiv.org/abs/hep-ph/9708393}{{\ttfamily hep-ph/9708393}}].

\bibitem{Grzadkowski:2018nbc}
B.~Grzadkowski and D.~Huang, \emph{{Spontaneous $CP$-Violating Electroweak
  Baryogenesis and Dark Matter from a Complex Singlet Scalar}},
  \href{https://doi.org/10.1007/JHEP08(2018)135}{\emph{JHEP} {\bfseries 08}
  (2018) 135} [\href{https://arxiv.org/abs/1807.06987}{{\ttfamily
  1807.06987}}].

\bibitem{Ellis:2019flb}
S.~A.~R. Ellis, S.~Ipek and G.~White, \emph{{Electroweak Baryogenesis from
  Temperature-Varying Couplings}},
  \href{https://arxiv.org/abs/1905.11994}{{\ttfamily 1905.11994}}.

\bibitem{Huang:2018aja}
F.~P. Huang, Z.~Qian and M.~Zhang, \emph{{Exploring dynamical CP violation
  induced baryogenesis by gravitational waves and colliders}},
  \href{https://doi.org/10.1103/PhysRevD.98.015014}{\emph{Phys. Rev.}
  {\bfseries D98} (2018) 015014}
  [\href{https://arxiv.org/abs/1804.06813}{{\ttfamily 1804.06813}}].

\bibitem{Abel:1996cr}
S.~A. Abel, \emph{{Destabilizing divergences in the NMSSM}},
  \href{https://doi.org/10.1016/S0550-3213(96)00470-1}{\emph{Nucl. Phys.}
  {\bfseries B480} (1996) 55}
  [\href{https://arxiv.org/abs/hep-ph/9609323}{{\ttfamily hep-ph/9609323}}].

\bibitem{Panagiotakopoulos:1998yw}
C.~Panagiotakopoulos and K.~Tamvakis, \emph{{Stabilized NMSSM without domain
  walls}}, \href{https://doi.org/10.1016/S0370-2693(98)01493-2}{\emph{Phys.
  Lett.} {\bfseries B446} (1999) 224}
  [\href{https://arxiv.org/abs/hep-ph/9809475}{{\ttfamily hep-ph/9809475}}].

\bibitem{Panagiotakopoulos:1999ah}
C.~Panagiotakopoulos and K.~Tamvakis, \emph{{New minimal extension of MSSM}},
  \href{https://doi.org/10.1016/S0370-2693(99)01247-2}{\emph{Phys. Lett.}
  {\bfseries B469} (1999) 145}
  [\href{https://arxiv.org/abs/hep-ph/9908351}{{\ttfamily hep-ph/9908351}}].

\bibitem{Elliott:1993ex}
T.~Elliott, S.~F. King and P.~L. White, \emph{{Supersymmetric Higgs bosons at
  the limit}}, \href{https://doi.org/10.1016/0370-2693(93)91107-X}{\emph{Phys.
  Lett.} {\bfseries B305} (1993) 71}
  [\href{https://arxiv.org/abs/hep-ph/9302202}{{\ttfamily hep-ph/9302202}}].

\bibitem{Elliott:1993uc}
T.~Elliott, S.~F. King and P.~L. White, \emph{{Squark contributions to Higgs
  boson masses in the next-to-minimal supersymmetric standard model}},
  \href{https://doi.org/10.1016/0370-2693(93)91321-D}{\emph{Phys. Lett.}
  {\bfseries B314} (1993) 56}
  [\href{https://arxiv.org/abs/hep-ph/9305282}{{\ttfamily hep-ph/9305282}}].

\bibitem{Elliott:1993bs}
T.~Elliott, S.~F. King and P.~L. White, \emph{{Radiative corrections to Higgs
  boson masses in the next-to-minimal supersymmetric Standard Model}},
  \href{https://doi.org/10.1103/PhysRevD.49.2435}{\emph{Phys. Rev.} {\bfseries
  D49} (1994) 2435} [\href{https://arxiv.org/abs/hep-ph/9308309}{{\ttfamily
  hep-ph/9308309}}].

\bibitem{Patel:2011th}
H.~H. Patel and M.~J. Ramsey-Musolf, \emph{{Baryon Washout, Electroweak Phase
  Transition, and Perturbation Theory}},
  \href{https://doi.org/10.1007/JHEP07(2011)029}{\emph{JHEP} {\bfseries 07}
  (2011) 029} [\href{https://arxiv.org/abs/1101.4665}{{\ttfamily 1101.4665}}].

\bibitem{Romao:1986jy}
J.~C. Romao, \emph{{Spontaneous {CP} Violation in {SUSY} Models: A No Go
  Theorem}}, \href{https://doi.org/10.1016/0370-2693(86)90522-8}{\emph{Phys.
  Lett.} {\bfseries B173} (1986) 309}.

\bibitem{Ferreira:2019iqb}
P.~M. Ferreira, M.~Mühlleitner, R.~Santos, G.~Weiglein and J.~Wittbrodt,
  \emph{{Vacuum Instabilities in the N2HDM}},
  \href{https://arxiv.org/abs/1905.10234}{{\ttfamily 1905.10234}}.

\bibitem{Nielsen:1975fs}
N.~K. Nielsen, \emph{{On the Gauge Dependence of Spontaneous Symmetry Breaking
  in Gauge Theories}},
  \href{https://doi.org/10.1016/0550-3213(75)90301-6}{\emph{Nucl. Phys.}
  {\bfseries B101} (1975) 173}.

\bibitem{DiLuzio:2014bua}
L.~Di~Luzio and L.~Mihaila, \emph{{On the gauge dependence of the Standard
  Model vacuum instability scale}},
  \href{https://doi.org/10.1007/JHEP06(2014)079}{\emph{JHEP} {\bfseries 06}
  (2014) 079} [\href{https://arxiv.org/abs/1404.7450}{{\ttfamily 1404.7450}}].

\bibitem{Laine:2017hdk}
M.~Laine, M.~Meyer and G.~Nardini, \emph{{Thermal phase transition with full
  2-loop effective potential}},
  \href{https://doi.org/10.1016/j.nuclphysb.2017.04.023}{\emph{Nucl. Phys.}
  {\bfseries B920} (2017) 565}
  [\href{https://arxiv.org/abs/1702.07479}{{\ttfamily 1702.07479}}].

\bibitem{Arnold:1992rz}
P.~B. Arnold and O.~Espinosa, \emph{{The Effective potential and first order
  phase transitions: Beyond leading-order}},
  \href{https://doi.org/10.1103/physrevd.50.6662.2,
  10.1103/PhysRevD.47.3546}{\emph{Phys. Rev.} {\bfseries D47} (1993) 3546}
  [\href{https://arxiv.org/abs/hep-ph/9212235}{{\ttfamily hep-ph/9212235}}].

\bibitem{Basler:2016obg}
P.~Basler, M.~Krause, M.~Muhlleitner, J.~Wittbrodt and A.~Wlotzka,
  \emph{{Strong First Order Electroweak Phase Transition in the CP-Conserving
  2HDM Revisited}}, \href{https://doi.org/10.1007/JHEP02(2017)121}{\emph{JHEP}
  {\bfseries 02} (2017) 121}
  [\href{https://arxiv.org/abs/1612.04086}{{\ttfamily 1612.04086}}].

\bibitem{Basler:2018cwe}
P.~Basler and M.~Mühlleitner, \emph{{BSMPT (Beyond the Standard Model Phase
  Transitions): A tool for the electroweak phase transition in extended Higgs
  sectors}}, \href{https://doi.org/10.1016/j.cpc.2018.11.006}{\emph{Comput.
  Phys. Commun.} {\bfseries 237} (2019) 62}
  [\href{https://arxiv.org/abs/1803.02846}{{\ttfamily 1803.02846}}].

\bibitem{Coleman:1977py}
S.~R. Coleman, \emph{{The Fate of the False Vacuum. 1. Semiclassical Theory}},
  \href{https://doi.org/10.1103/PhysRevD.15.2929,
  10.1103/PhysRevD.16.1248}{\emph{Phys. Rev.} {\bfseries D15} (1977) 2929}.

\bibitem{Callan:1977pt}
C.~G. Callan, Jr. and S.~R. Coleman, \emph{{The Fate of the False Vacuum. 2.
  First Quantum Corrections}},
  \href{https://doi.org/10.1103/PhysRevD.16.1762}{\emph{Phys. Rev.} {\bfseries
  D16} (1977) 1762}.

\bibitem{Linde:1980tt}
A.~D. Linde, \emph{{Fate of the False Vacuum at Finite Temperature: Theory and
  Applications}},
  \href{https://doi.org/10.1016/0370-2693(81)90281-1}{\emph{Phys. Lett.}
  {\bfseries 100B} (1981) 37}.

\bibitem{Wainwright:CosmoTransition}
C.~L. Wainwright, \emph{{CosmoTransitions: Computing Cosmological Phase
  Transition Temperatures and Bubble Profiles with Multiple Fields}},
  \href{https://doi.org/10.1016/j.cpc.2012.04.004}{\emph{Comput. Phys. Commun.}
  {\bfseries 183} (2012) 2006}
  [\href{https://arxiv.org/abs/1109.4189}{{\ttfamily 1109.4189}}].

\bibitem{McLerran:1990zh}
L.~D. McLerran, M.~E. Shaposhnikov, N.~Turok and M.~B. Voloshin, \emph{{Why the
  baryon asymmetry of the universe is approximately $10^{-10}$}},
  \href{https://doi.org/10.1016/0370-2693(91)91794-V}{\emph{Phys. Lett.}
  {\bfseries B256} (1991) 451}.

\bibitem{Dine:1991ck}
M.~Dine, P.~Huet and R.~L. Singleton, Jr., \emph{{Baryogenesis at the
  electroweak scale}},
  \href{https://doi.org/10.1016/0550-3213(92)90113-P}{\emph{Nucl. Phys.}
  {\bfseries B375} (1992) 625}.

\bibitem{Athron:2014yba}
P.~Athron, J.-h. Park, D.~Stöckinger and A.~Voigt, \emph{{FlexibleSUSY -- A
  spectrum generator generator for supersymmetric models}},
  \href{https://doi.org/10.1016/j.cpc.2014.12.020}{\emph{Comput. Phys. Commun.}
  {\bfseries 190} (2015) 139}
  [\href{https://arxiv.org/abs/1406.2319}{{\ttfamily 1406.2319}}].

\bibitem{Athron:2017fvs}
P.~Athron, M.~Bach, D.~Harries, T.~Kwasnitza, J.-h. Park, D.~Stöckinger
  et~al., \emph{{FlexibleSUSY 2.0: Extensions to investigate the phenomenology
  of SUSY and non-SUSY models}},
  \href{https://arxiv.org/abs/1710.03760}{{\ttfamily 1710.03760}}.

\bibitem{Allanach:2001kg}
B.~Allanach, \emph{{SOFTSUSY: a program for calculating supersymmetric
  spectra}},
  \href{https://doi.org/10.1016/S0010-4655(01)00460-X}{\emph{Comput.Phys.Commun.}
  {\bfseries 143} (2002) 305}
  [\href{https://arxiv.org/abs/hep-ph/0104145}{{\ttfamily hep-ph/0104145}}].

\bibitem{Allanach:2013kza}
B.~Allanach, P.~Athron, L.~C. Tunstall, A.~Voigt and A.~Williams,
  \emph{{Next-to-Minimal SOFTSUSY}},
  \href{https://doi.org/10.1016/j.cpc.2014.04.015}{\emph{Comput.Phys.Commun.}
  {\bfseries 185} (2014) 2322}
  [\href{https://arxiv.org/abs/1311.7659}{{\ttfamily 1311.7659}}].

\bibitem{Staub:2009bi}
F.~Staub, \emph{{From Superpotential to Model Files for FeynArts and
  CalcHep/CompHep}},
  \href{https://doi.org/10.1016/j.cpc.2010.01.011}{\emph{Comput.Phys.Commun.}
  {\bfseries 181} (2010) 1077}
  [\href{https://arxiv.org/abs/0909.2863}{{\ttfamily 0909.2863}}].

\bibitem{Staub:2010jh}
F.~Staub, \emph{{Automatic Calculation of supersymmetric Renormalization Group
  Equations and Self Energies}},
  \href{https://doi.org/10.1016/j.cpc.2010.11.030}{\emph{Comput.Phys.Commun.}
  {\bfseries 182} (2011) 808}
  [\href{https://arxiv.org/abs/1002.0840}{{\ttfamily 1002.0840}}].

\bibitem{Staub:2012pb}
F.~Staub, \emph{{SARAH 3.2: Dirac Gauginos, UFO output, and more}},
  \href{https://doi.org/10.1016/j.cpc.2013.02.019}{\emph{Computer Physics
  Communications} {\bfseries 184} (2013) pp. 1792}
  [\href{https://arxiv.org/abs/1207.0906}{{\ttfamily 1207.0906}}].

\bibitem{Staub:2013tta}
F.~Staub, \emph{{SARAH 4: A tool for (not only SUSY) model builders}},
  \href{https://doi.org/10.1016/j.cpc.2014.02.018}{\emph{Comput.Phys.Commun.}
  {\bfseries 185} (2014) 1773}
  [\href{https://arxiv.org/abs/1309.7223}{{\ttfamily 1309.7223}}].

\bibitem{Fowlie:2018eiu}
A.~Fowlie, \emph{{A fast C++ implementation of thermal functions}},
  \href{https://doi.org/10.1016/j.cpc.2018.02.015}{\emph{Comput. Phys. Commun.}
  {\bfseries 228} (2018) 264}
  [\href{https://arxiv.org/abs/1802.02720}{{\ttfamily 1802.02720}}].

\bibitem{Staub:2016dxq}
F.~Staub et~al., \emph{{Precision tools and models to narrow in on the 750 GeV
  diphoton resonance}},
  \href{https://doi.org/10.1140/epjc/s10052-016-4349-5}{\emph{Eur. Phys. J.}
  {\bfseries C76} (2016) 516}
  [\href{https://arxiv.org/abs/1602.05581}{{\ttfamily 1602.05581}}].

\bibitem{Bernon:2015hsa}
J.~Bernon and B.~Dumont, \emph{{Lilith: a tool for constraining new physics
  from Higgs measurements}},
  \href{https://doi.org/10.1140/epjc/s10052-015-3645-9}{\emph{Eur. Phys. J.}
  {\bfseries C75} (2015) 440}
  [\href{https://arxiv.org/abs/1502.04138}{{\ttfamily 1502.04138}}].

\bibitem{Feroz:2007kg}
F.~Feroz and M.~P. Hobson, \emph{{Multimodal nested sampling: an efficient and
  robust alternative to MCMC methods for astronomical data analysis}},
  \href{https://doi.org/10.1111/j.1365-2966.2007.12353.x}{\emph{Mon. Not. Roy.
  Astron. Soc.} {\bfseries 384} (2008) 449}
  [\href{https://arxiv.org/abs/0704.3704}{{\ttfamily 0704.3704}}].

\bibitem{Feroz:2008xx}
F.~Feroz, M.~P. Hobson and M.~Bridges, \emph{{MultiNest: an efficient and
  robust Bayesian inference tool for cosmology and particle physics}},
  \href{https://doi.org/10.1111/j.1365-2966.2009.14548.x}{\emph{Mon. Not. Roy.
  Astron. Soc.} {\bfseries 398} (2009) 1601}
  [\href{https://arxiv.org/abs/0809.3437}{{\ttfamily 0809.3437}}].

\bibitem{2013arXiv1306.2144F}
F.~{Feroz}, M.~P. {Hobson}, E.~{Cameron} and A.~N. {Pettitt}, \emph{{Importance
  Nested Sampling and the MultiNest Algorithm}},
  \href{https://arxiv.org/abs/1306.2144}{{\ttfamily 1306.2144}}.

\bibitem{arXiv:0811.4169}
P.~Bechtle, O.~Brein, S.~Heinemeyer, G.~Weiglein and K.~E. Williams,
  \emph{{HiggsBounds: Confronting Arbitrary Higgs Sectors with Exclusion Bounds
  from LEP and the Tevatron}},
  \href{https://doi.org/10.1016/j.cpc.2009.09.003}{\emph{Comput. Phys. Commun.}
  {\bfseries 181} (2010) 138}
  [\href{https://arxiv.org/abs/0811.4169}{{\ttfamily 0811.4169}}].

\bibitem{arXiv:1102.1898}
P.~Bechtle, O.~Brein, S.~Heinemeyer, G.~Weiglein and K.~E. Williams,
  \emph{{HiggsBounds 2.0.0: Confronting Neutral and Charged Higgs Sector
  Predictions with Exclusion Bounds from LEP and the Tevatron}},
  \href{https://doi.org/10.1016/j.cpc.2011.07.015}{\emph{Comput. Phys. Commun.}
  {\bfseries 182} (2011) 2605}
  [\href{https://arxiv.org/abs/1102.1898}{{\ttfamily 1102.1898}}].

\bibitem{arXiv:1301.2345}
P.~Bechtle et~al., \emph{{Recent Developments in HiggsBounds and a Preview of
  HiggsSignals}}, {\emph{PoS} {\bfseries CHARGED2012} (2012) 024}
  [\href{https://arxiv.org/abs/1301.2345}{{\ttfamily 1301.2345}}].

\bibitem{arXiv:1311.0055}
P.~Bechtle et~al., \emph{{HiggsBounds-4: Improved Tests of Extended Higgs
  Sectors against Exclusion Bounds from LEP, the Tevatron and the LHC}},
  {\emph{Eur. Phys. J.} {\bfseries C74} (2014) 2693}
  [\href{https://arxiv.org/abs/1311.0055}{{\ttfamily 1311.0055}}].

\bibitem{arXiv:1507.06706}
P.~Bechtle, S.~Heinemeyer, O.~Stal, T.~Stefaniak and G.~Weiglein,
  \emph{{Applying Exclusion Likelihoods from LHC Searches to Extended Higgs
  Sectors}},  \href{https://arxiv.org/abs/1507.06706}{{\ttfamily 1507.06706}}.

\bibitem{Baglio:2013iia}
J.~Baglio, R.~Gröber, M.~Mühlleitner, D.~T. Nhung, H.~Rzehak, M.~Spira
  et~al., \emph{{NMSSMCALC: A Program Package for the Calculation of
  Loop-Corrected Higgs Boson Masses and Decay Widths in the (Complex) NMSSM}},
  \href{https://doi.org/10.1016/j.cpc.2014.08.005}{\emph{Comput. Phys. Commun.}
  {\bfseries 185} (2014) 3372}
  [\href{https://arxiv.org/abs/1312.4788}{{\ttfamily 1312.4788}}].

\bibitem{Aaboud:2017ayj}
{\scshape ATLAS} collaboration, \emph{{Search for a scalar partner of the top
  quark in the jets plus missing transverse momentum final state at
  $\sqrt{s}$=13 TeV with the ATLAS detector}},
  \href{https://doi.org/10.1007/JHEP12(2017)085}{\emph{JHEP} {\bfseries 12}
  (2017) 085} [\href{https://arxiv.org/abs/1709.04183}{{\ttfamily
  1709.04183}}].

\bibitem{CMS:2019qkm}
{\scshape CMS} collaboration, \emph{{Search for direct top squark pair
  production in events with one lepton, jets and missing transverse energy at
  13 TeV}},  tech. rep., 2019.
\newblock \href{https://cds.cern.ch/record/2682157}{CMS-PAS-SUS-19-009}.

\bibitem{Sirunyan:2018qlb}
{\scshape CMS} collaboration, \emph{{Search for a new scalar resonance decaying
  to a pair of Z bosons in proton-proton collisions at $\sqrt{s}=13 $ TeV}},
  \href{https://doi.org/10.1007/JHEP06(2018)127}{\emph{JHEP} {\bfseries 06}
  (2018) 127} [\href{https://arxiv.org/abs/1804.01939}{{\ttfamily
  1804.01939}}].

\bibitem{AbdusSalam:2017uzr}
S.~AbdusSalam, \emph{{Testing Higgs boson scenarios in the phenomenological
  NMSSM}}, \href{https://doi.org/10.1140/epjc/s10052-019-6953-7}{\emph{Eur.
  Phys. J.} {\bfseries C79} (2019) 442}
  [\href{https://arxiv.org/abs/1710.10785}{{\ttfamily 1710.10785}}].

\bibitem{Vieu:2018zze}
T.~Vieu, A.~P. Morais and R.~Pasechnik, \emph{{Multi-peaked signatures of
  primordial gravitational waves from multi-step electroweak phase
  transition}},  \href{https://arxiv.org/abs/1802.10109}{{\ttfamily
  1802.10109}}.

\bibitem{Simplex}
J.~Nelder and R.~Mead, \emph{{A Simplex Method for Function Minimization}},
  \href{https://doi.org/10.1093/comjnl/7.4.308}{\emph{Comput.J.} {\bfseries 7}
  (1965) 308}.

\end{thebibliography}\endgroup

\end{document}